%% file: ms.tex
\documentclass[12pt,preprint]{aastex}

\slugcomment{To be submitted to Astrophysical Journal}

\shorttitle{Dust and SN~2002hh}
\shortauthors{Meikle et al.}

\begin{document}


\title{A Spitzer Space Telescope study of SN~2002hh: an infrared echo
from a Type~IIP supernova}


\author{W. P. S. Meikle,\altaffilmark{1}
S. Mattila,\altaffilmark{2} 
C. L. Gerardy,\altaffilmark{1} 
R. Kotak,\altaffilmark{3} 
M. Pozzo,\altaffilmark{1} 
S. D. van Dyk,\altaffilmark{4} 
D. Farrah,\altaffilmark{5} 
R. A. Fesen,\altaffilmark{6} 
A. V. Filippenko,\altaffilmark{7} 
C. Fransson,\altaffilmark{8} 
P. Lundqvist,\altaffilmark{8} 
J. Sollerman,\altaffilmark{9} 
J. C. Wheeler\altaffilmark{10}} 

\altaffiltext{1}{Astrophysics Group, Blackett Laboratory, Imperial
College London, Prince Consort Road, London SW7 2AZ, United Kingdom;
p.meikle@imperial.ac.uk, c.gerardy@imperial.ac.uk,
m.pozzo@imperial.ac.uk}
\altaffiltext{2}{Department of Physics and Astronomy, Queen's
University Belfast, County Antrim, BT7 1NN, United Kingdom;
s.mattila@qub.ac.uk}
\altaffiltext{3}{European Southern Observatory,
Karl-Schwarzschild-Str. 2, D-85748 Garching bei München, Germany;
rkotak@eso.org}
\altaffiltext{4}{Infrared Processing and Analysis Center, California
Institute of Technology, Mail Code 100-22, 770 South Wilson Avenue,
Pasadena, CA 91125; vandyk@ipac.caltech.edu}
\altaffiltext{5}{Department of Astronomy, 106 Space Sciences Building,
Cornell University, Ithaca, NY 14853; duncan@isc.astro.cornell.edu}
\altaffiltext{6}{Department of Physics and Astronomy, Dartmouth
College, 6127 Wilder Laboratory, Hanover, NH 03755-3528;
fesen@snr.dartmouth.edu}
\altaffiltext{7}{Dept. of Astronomy, 601 Campbell Hall, University of
California, Berkeley, CA 94720-3411; alex@astro.berkeley.edu}
\altaffiltext{8}{Stockholm University, AlbaNova University Center,
Stockholm Observatory, Department of Astronomy, SE-106 91 Stockholm,
Sweden; claes@astro.su.se, peter@astro.su.se}
\altaffiltext{9}{Dark Cosmology Centre, Niels Bohr Institute,
University of Copenhagen, Juliane Maries Vej 30, 2100 Copenhagen,
Denmark; jesper@astro.ku.dk}
\altaffiltext{10}{The University of Texas at Austin, Department of
Astronomy, Austin, TX 78712; wheel@astro.as.utexas.edu}




\begin{abstract}
We present late-time (590--994~d) mid-IR photometry of the normal, but
highly-reddened Type~IIP supernova SN~2002hh.  Bright, cool,
slowly-fading emission is detected from the direction of the
supernova. Most of this flux appears not to be driven by the supernova
event but instead probably originates in a cool, obscured
star-formation region or molecular cloud along the line-of-sight.  We
also show, however, that the declining component of the flux is
consistent with an SN-powered IR~echo from a dusty progenitor CSM.
Mid-IR emission could also be coming from newly-condensed dust and/or
an ejecta/CSM impact but their contributions are likely to be small.
For the case of a CSM-IR~echo, we infer a dust mass of as little as
0.036~M$_{\odot}$ with a corresponding CSM mass of
$3.6(0.01/r_{dg})$~M$_{\odot}$ where $r_{dg}$ is the dust-to-gas mass
ratio.  Such a CSM would have resulted from episodic mass loss whose
rate declined significantly about 28,000~years ago.  Alternatively, an
IR~echo from a surrounding, dense, dusty molecular cloud might also
have been responsible for the fading component. Either way, this is
the first time that an IR~echo has been clearly identified in a
Type~IIP supernova.  We find no evidence for or against the proposal
that Type~IIP supernovae produce large amounts of dust via grain
condensation in the ejecta. However, within the CSM-IR~echo scenario,
the mass of dust derived implies that the {\it progenitors} of the
most common of core-collapse supernovae may make an important
contribution to the universal dust content.
\end{abstract}





\section{Introduction}

A major goal in the study of core-collapse supernovae (CCSNe) is to
test the proposal that they are, or have been, a significant source of
dust in the universe.  The physical conditions believed to prevail in
the expanding SN ejecta make this an attractive idea. Large abundances
of suitable refractory elements are present. Cooling by adiabatic
expansion and molecular emission takes place.  Dynamical instabilities
can produce density enhancements or ``clumping''.  Support for these
ideas is provided by isotopic anomalies in meteorites which indicate
that some grains must have formed in CCSNe \citep{cla97}.  Interest in
CCSNe as dust producers has increased recently due the problem of
accounting for the presence of dust at high redshifts
\citep{fal89,fal96,pei91,pet97,ber03}.  In these early eras, much less
dust production from novae and AGB stars is expected since fewer stars
will have evolved past the main-sequence phase.  Consequently,
supernovae arising from Population~III stars are proposed as the main
early-universe source of dust.  Models of dust formation in supernovae
\citep{tod01,noz03} succeed in producing copious amounts of dust -
around 0.1-1 M$_{\odot}$ even in the low-metallicity environments at
high redshifts.  This is easily enough to make supernovae a major
contributor to the dust content of the universe.  \\

Observations of SN~1987A yielded indirect evidence that large masses
of dust can be produced in CCSNe.  Around 500~days post-explosion, its
UVOIR light curve dipped below the radioactive deposition input
\citep{whi89}. At the same time the mid-IR flux increased, nicely
accounting for the UVOIR deficit (\citet{whi89} and refs. therein).
This suggests substantial dust formation in the ejecta, although the
actual mass involved is uncertain.  Spectroscopic observations of
SN~1987A \citep{dan91,dwe92,spy92} indicate that, at 600--700~d, most
of the silicon and iron emission disappeared from the optical/near-IR
wavelength range.  If we attribute this loss to depletion onto grains
(e.g., Mg$_2$SiO$_4$, MgSiO$_3$, Fe$_3$O$_4$) then this implies a
large mass of dust - perhaps as much as 0.1~M$_{\odot}$.  However, it
may be that cooling and expansion to below the critical density can
account for much of the fading of the spectral lines
\citep{fra87,luc91,mei93}.\\

In spite of the SN-grain condensation hypothesis being over 30 years
old \citep{cer67,hoy70} there is still no {\it direct} observational
evidence that SNe are major dust sources. In particular, it is not
known if ordinary Type~IIP SNe produce large amounts of dust.  Dust
condensation in CCSN ejecta can be observed directly in two ways.  The
first method makes use of the attenuating effect on the red wings of
the broad ejecta lines during the nebular phase.  This technique has
the advantage of being relatively unambiguous in its ability to
demonstrate the presence of new dust, although extraction of
quantitative information about the quantity and nature of the grains
may be more difficult. The method is technically challenging in that
it requires high S/N, high spectral resolution observations taken at
as much as 2~years post-explosion.  Consequently, only three CCSNe -
the Type~IIpec SN~1987A \citep{dan91}, the Type~IIn SN~1998S
\citep{poz04} and the Type IIP SN~1999em \citep{elm03} - have been
studied in this way.  In all three cases, dust masses of just
10$^{-4}$--10$^{-3}$~M$_{\odot}$ were found to be adequate to account
for the attenuation, although the actual mass of dust could be larger.
In SNe~1987A and 1998S, variation of the strength of the attenuation
from optical to NIR wavelengths was much weaker than would be expected
from an optically-thin screen of small grains.  Thus, either the grain
size was remarkably large ($>$5~$\mu$m radius) or the grains existed
in optically thick clumps.  The former explanation is unlikely since
such large grains are not produced in grain condensation models
\citep{tod01}. On the other hand, as mentioned above, clumping due to
dynamical instabilities is quite plausible.  The mass of dust may also
have been underestimated if not all of it is ``backlit'' by spectral
line emission. The Elmhamdi et al. result for SN~1999em is of special
importance since this supernova was a normal Type~IIP event.  Prior to
the launch of the Spitzer Space Telescope ({\it Spitzer}), this stood
as the only direct observational evidence of dust formation in a
typical CCSN. \\

The other way of directly studying newly-condensed dust is by means of
thermal IR emission from the grains.  Thirteen CCSNe have exhibited
late-time NIR excesses (\citet{ger02} and refs. therein).  (Note: The
peculiar Type~Ia SN~2002ic has also shown a strong IR excess
\citet{kot04}.)  This is attributed to thermal emission from hot
grains. Curiously, 12 of the CCSNe were not of Type~IIP, while for the
thirteenth, SN~1982L, the subtype is unknown.  Moreover, it cannot be
assumed that the NIR excess was due to emission from newly-formed
grains. An alternative mechanism is thermal emission from pre-existing
dust in the progenitor wind, heated by the supernova peak luminosity -
the so-called ``IR~echo''. Heating of CSM dust by ejecta/shock
interaction is another possibility.  Only for SNe~1987A and 1998S has
persuasive IR-excess-based evidence for dust condensation during an
observed supernova explosion been put forward
\citep{luc91,dwe92,mei93,roc93,woo93,poz04}. In both objects it was
found that a dust mass of only $\sim$10$^{-3}$~M$_{\odot}$ was
sufficient to account for the observed IR flux, although a
considerably larger mass could have been concealed in optically thick
clumps.  Mid-IR studies of the supernova remnant Cassiopeia~A
\citep{dwe87,lag96,dou01} indicate that dust formation took place
during its explosion, but again the mass of directly-observed dust is
small.  The claim by \citet{dun03} that at least 2~M$_\odot$ of dust
formed in the Cas~A supernova has been contested by \citet{dwe04} and
\citet{kra04}.  \\

The availability of {\it Spitzer} has provided an excellent
opportunity for us to test the dust-condensation hypothesis in a
statistically significant number of typical supernovae.  It provides
high-sensitivity imaging over the mid-IR covering the likely peak of
the dust thermal emission spectrum. This can provide a superior
measure of the total flux, temperature and, possibly, dust emissivity
than can be achieved at shorter wavelengths.  Moreover, the longer
wavelength coverage of {\it Spitzer} lets us detect cooler grains, and
see more deeply into dust clumps than was previously possible for
typical nearby CCSNe.  In addition, multi-epoch observations with {\it
Spitzer} may allow us to distinguish between dust condensation and
IR~echoes via the light curve shape e.g. a flat-topped light curve
favours an IR~echo (e.g. \citet{dwe83}). \\

SN~2002hh was discovered in NGC~6946 on UT 2002 October 31.1
\citep{li02}.  The supernova was not detected on a KAIT image on
October 26.1 at a limiting magnitude of $\sim$19.0. We therefore adopt
2002 October 29$\pm$2 (JD 2452577$\pm$2) as the explosion epoch.
NGC~6946 is at a distance of 5.9 Mpc \citep{kar00} and has produced 7
other SNe.  From optical spectra taken on 2002 November 2
\citet{fil02} identified SN~2002hh as a young, highly reddened Type-II
supernova.  It peaked at V$\sim$17.2 on 2002 November 5.  A detailed
study of the optical and near-IR evolution of SN~2002hh during its
first year has been presented by \citet{poz06}. Their light curves
show SN~2002hh to have been a Type~IIP event.  Using $JK_s$ images
taken on 2002 Nov. 18.86 , \citet{mei02} inferred a host-galaxy
extinction $A_V$ of $\sim$5.0. This was confirmed by \citet{poz06},
who propose a two-component extinction law to account for the spectral
reddening and the depth of K~I interstellar absorption. \\ 

\citet{sto02} detected radio emission at 8.435 and 22.485 GHz at
17~d. They suggested that ``the apparently optically thin character of
the radio emission, if confirmed, may indicate that the circumstellar
interaction is weak and is evolving unusually rapidly.''  SN 2002hh
was also detected at 1396.75 MHz by \citet{cha03} at 59~d.
\citet{poo02} detected X-ray emission at 27~d.  They state that the
rather hard, highly absorbed spectrum is as might be expected given
the high reddening, and that the low luminosity supports the radio
indication that little circumstellar interaction was taking place at
that time.  \citet{che06} find that an optical depth of unity occurred
at day~62 at 1.4 GHz, implying a mass loss rate of $7\times10^{-6}
(T_{cs}/10^5 K)^{3/4}v_{w1}$~M$_{\odot}$~yr$^{-1}$ where $T_{cs}$ is
the temperature of the CSM and $v_{w1}$ is the progenitor wind
velocity in units of 10~km~s$^{-1}$.  \citet{bes05} measured radio
emission at 4.860 GHz at 381~d, and at 1.425 GHz at 899~d.  Late-time
mid-IR observations have been reported by \citet{bar05} and these are
discussed below. \\

Here we present results of {\it Spitzer} observations of
SN~2002hh. This forms part of the work of our Mid-Infrared Supernova
Consortium (MISC) which is using {\it Spitzer} to observe a
substantial sample of nearby SNe of all types and over a wide range of
epochs.  Preliminary results from our SN~2002hh {\it Spitzer}
observations have been presented by \citet{mei05}. \\

\section{Observations}
SN~2002hh was observed with the Infrared Array Camera (IRAC)
\citep{faz04} on {\it Spitzer} in all four channels (3.6~$\mu$m,
4.5~$\mu$m, 5.8~$\mu$m, 8.0~$\mu$m) and at four epochs spanning 590 to
994~days post-explosion. The observation log is given in Table
\ref{tab1}.  As part of our {\it Spitzer} supernova programs
(PID:3248, 20256) the supernova was observed at epochs 684~d, 758~d
and 994~d.  For each filter, a full-array 20-point medium dither was
used.  The integration time per frame was 26.8~s, which kept the
counts per frame on the target at well below the non-linearity limit.
Thus, the total integration time per channel at a given epoch was
536~s.  The target was also observed serendipitously within the {\it
Spitzer Infrared Nearby Galaxies Survey (SINGS)} \citep{ken03} at
epochs 590~d and 758~d. In this case, for each filter a mapping scheme
in high-dynamic-range mode was used.  The total integration time per
channel at a given epoch was 214~s.  All the data were reprocessed in
the S11.0.2 pipeline.  We use the post-basic calibrated data (PBCD)
product throughout.  From the point of view of dust condensation, the
epochs covered are particularly interesting since, in the case of
SN~1987A, the mid-IR luminosity from the ejecta dust peaked around
600~d \citep{bou93}.\\

\section{Results}
\begin{figure}
\epsscale{1.0}
\plotone{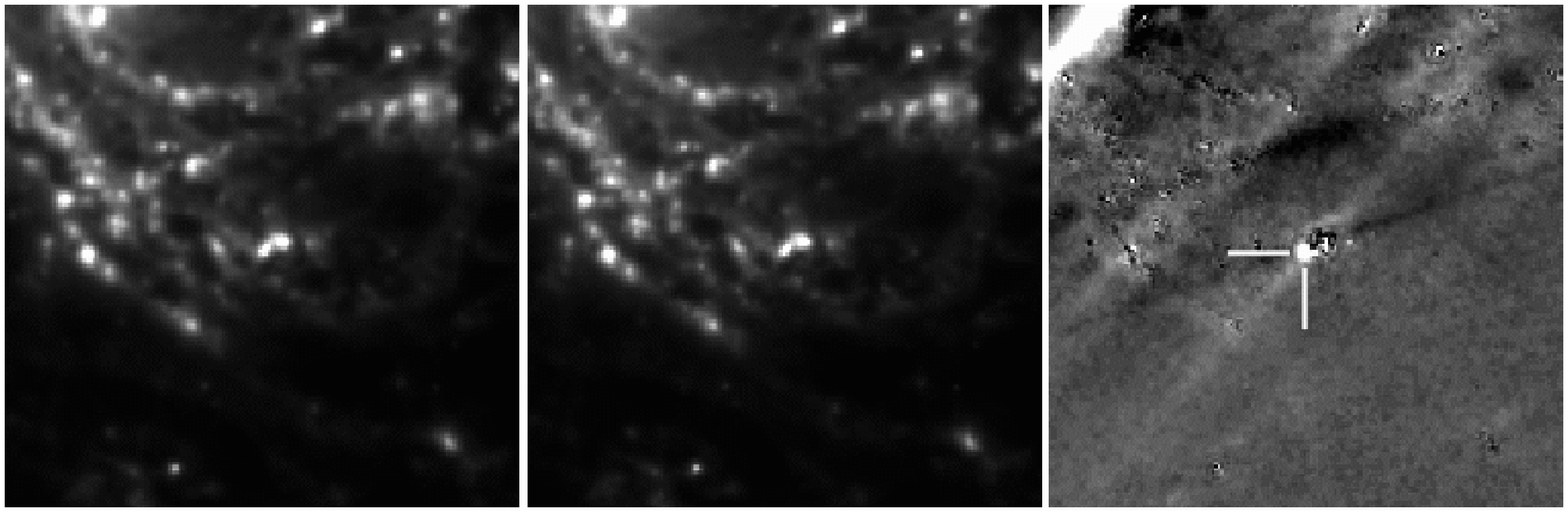}
\caption[]{ {\it Left, Centre} IRAC 8~$\mu$m images ($200\arcsec
\times 200\arcsec$ sections) of the SN~2002hh field taken,
respectively, at 590~d and 994~d post-explosion.  North is up and East
is to the left.  The centre of the field is dominated by a bright
2MASS star.  Extended emission to the approximate south-east of the
star is also apparent. There appears to be little difference in the
flux at the supernova position between the two images.  \\ {\it Right}
Subtraction of the centre image (994~d) from the left hand image
(590~d) after PSF and intensity matching.  The supernova, indicated by
the tick marks, is clearly visible.
\label{fig1}
}
\end{figure}

In Figure \ref{fig1} we show {\it Spitzer}~(IRAC) 8~$\mu$m images
($200\arcsec \times 200\arcsec$ sections) of the SN~2002hh field taken
at the earliest and latest epochs viz. 590~d (left-hand image) and
994~d (centre image). The supernova is located within a complex
distribution of sources forming part of the spiral arm structure of
NGC~6946. The field is dominated by a bright star lying about
9\arcsec~ from the supernova at P.A. 298~deg.  The star's magnitudes
are $B=16.1$, $V=15.2$ (NED).  This star is also recorded in the 2MASS
catalogue as J20344320+6007234, with magnitudes $J=15.9$, $H=15.5$,
$K=15.3$.  (Henceforth, the star will be referred to as the ``2MASS
star''.)  Extended emission lies between the 2MASS star and the
supernova location.  We note that there is little apparent difference
between the images from one epoch to the next. The other three
channels show a similar lack of change.  However, the right-hand image
shows the effect of subtracting the 994~d image from the 590~d one
(details below). A bright point source is apparent at the supernova
location. This indicates the presence of a declining source and we
suggest below that the supernova is the cause of this. The
contribution which the supernova makes to the total flux from this
location is also discussed below.  \\

\begin{figure}
\epsscale{1.0}
\plotone{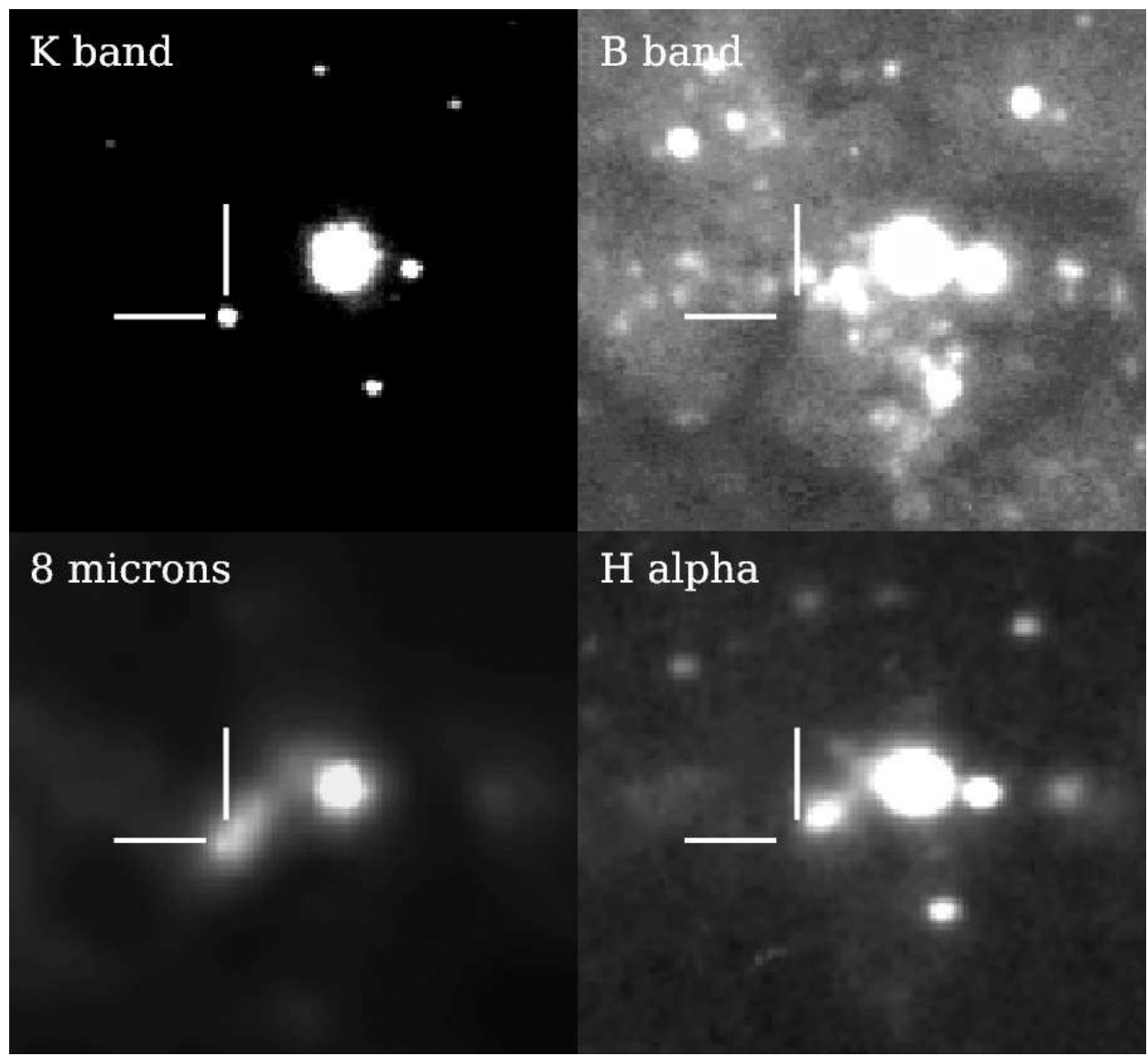}
\caption[]{The SN~2002hh field viewed at a variety of wavelengths and
epochs.  The FOV of each image is $38\arcsec \times 35\arcsec$. North
is up and East to the left.  All the images have been aligned
(including rebinning, rotation, and shifting) to the H$\alpha$
image. The SN location is marked with ticks in each image based on the
coordinates measured from the $K$~band image. In each image, the
bright 2MASS star lies about 9\arcsec~ and P.A. 298~deg from the
supernova location. The $K$~band image of SN~2002hh was obtained with
SPEX at the IRTF at 266~d \citep{poz06}, and the 8~$\mu$m {\it
Spitzer}~(IRAC) image was acquired at 590~d. The pre-explosion $B$~band
image is from \citet{lar99}.  Note the dust lane close to the
supernova position. The pre-explosion H$\alpha$ image is from
\citet{kna04}.  It illustrates the extended HII region close to the
supernova position.
\label{fig2}
}
\end{figure}

To illustrate the complexity of the SN~2002hh field, in Figure \ref{fig2} we
compare the 590~d 8~$\mu$m image with the same field at a number of
other wavelengths and epochs.  In a $K$~band 266~d image obtained at
the IRTF \citep{poz06} the supernova is clearly present.  A
pre-explosion $B$~band image \citep{lar99} indicates a dark dust-lane
close to the supernova location, while a pre-explosion H$\alpha$ image
\citep{kna04} reveals emission which extends from the 2MASS star
almost to the SN position. \\

Aperture photometry of the supernova was carried out on the PBCD data,
using the Starlink package {\sc gaia} \citep{dra02}.  The brightness
and complexity of the sources in the supernova field together with the
IRAC spatial resolution of FWHM $\sim$ 2\arcsec~ (at 8.0~$\mu$m) makes
the extraction of reliable photometry quite challenging.  We performed
simple aperture photometry on three different forms of the data viz.
(1) original images, (2) all images aligned to the same orientation
and position, and (3) PSF-matched, intensity-matched, subtracted
images.  A circular aperture of radius 3\farcs3 (2.7 pixel) was used
for the photometry. This is sufficiently large to encompass the flux
out to beyond the first Airy ring in the longest wavelength band
(8.0~$\mu$m) thus reducing the size of aperture correction needed in
the final flux determination. The aperture radius corresponds to a
distance of $\sim$100~pc at SN~2002hh.  A larger aperture was not used
in order to minimise contamination from the 2MASS star.  For each
measurement, the aperture was centred to within $\pm$0\farcs1 (0.08
pixels) of the supernova position as reported by \citet{li02} and
given in the NED and SIMBAD databases.  To test the accuracy of our
positioning procedure, we measured the WCS co-ordinates of the
centroid of isolated field stars for all the images.  The dispersion
in the 3.6 and 4.5~$\mu$m images was found to be 0\farcs18 in both
axes. For the 5.8 and 8~$\mu$m images the dispersion was 0\farcs34.
We note that the radio position of SN~2002hh reported by \citet{sto02}
differs from that given by \citet{li02} by 0\farcs5, or about 0.4 of
an IRAC pixel. We checked the effect of centering on the Stockdale et
al. position and found that it increased the measured fluxes by just
1--6\%. Aperture correction factors were applied as specified in
Table~5.7 of the {\it IRAC Data Handbook} \citep{rea06}.  For the
unsubtracted images, the sky was measured using a clipped mean sky
estimator and concentric annuli with radii chosen to avoid background
sources as far as possible.  Measurements were obtained for two
annulus settings viz. inner/outer radius values of 4/5 and 7/8 times
the aperture radius, respectively, and the mean flux value was adopted
for each channel/epoch.  For the 758~d flux, the mean of the SINGS and
MISC data was adopted.  The error contribution caused by position
uncertainty was estimated by offsetting the aperture by an amount
equal to the position uncertainty (dispersion) described above and
re-measuring the flux. This was carried out for a number of directions
from the WCS co-ordinates of the supernova.\\

\begin{figure}
\epsscale{1.0}
\plotone{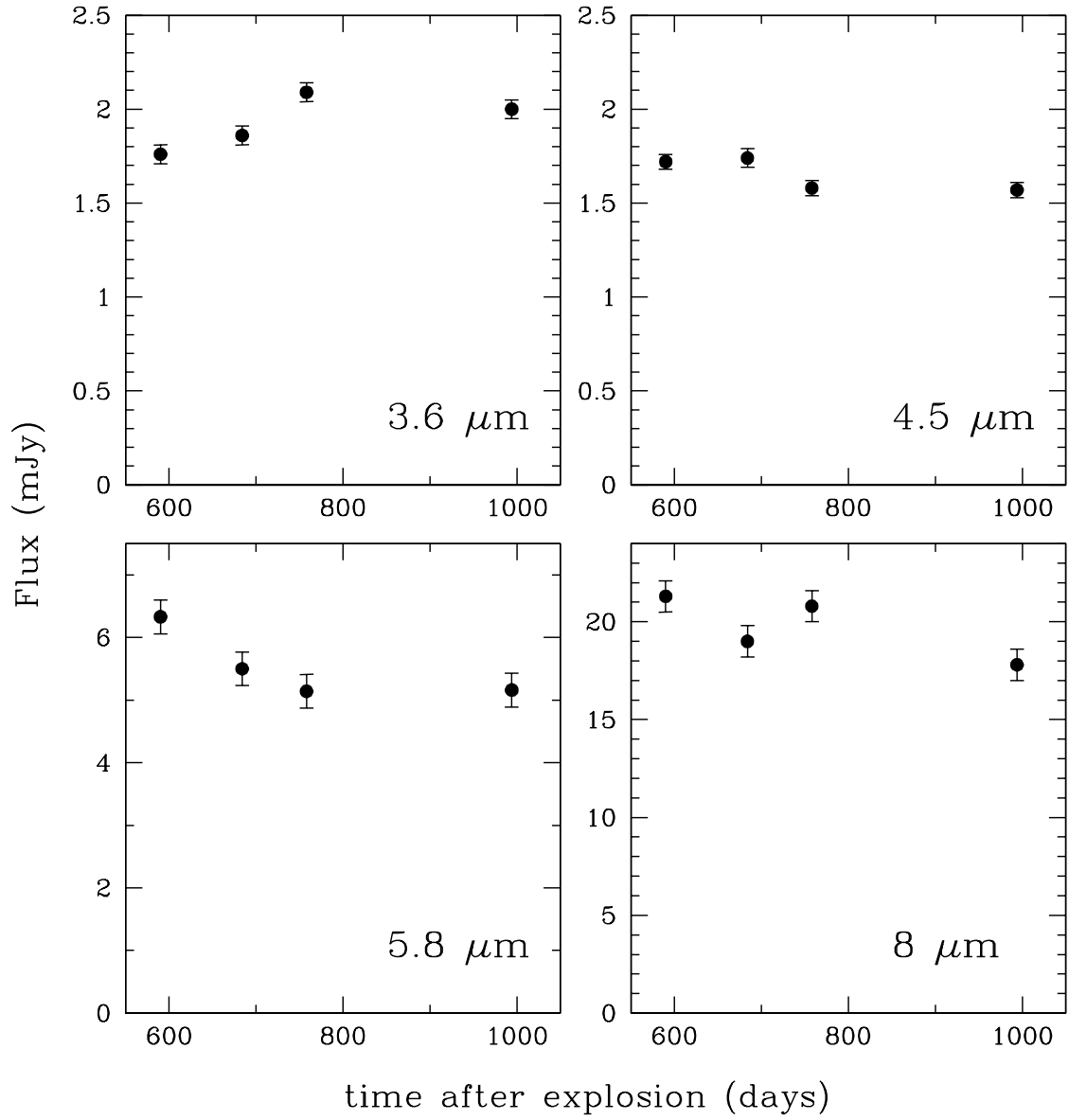}
\caption[]{ Mid-IR photometry of the SN~2002hh region using a
  6\farcs5 diameter aperture centred at the supernova position in
  aligned, unsubtracted images.
\label{fig3}
}
\end{figure}

Aperture photometry of the supernova region using the original
unaligned, unsubtracted images reveals only a hint of temporal
variation between 590~d and 994~d (Table \ref{tab2}).  While the flux at
3.6~$\mu$m exhibits an increase of about 0.2~mJy, declines of about
0.2~mJy, 1~mJy and 3~mJy appear to have occurred at 4.5, 5.8 and
8.0~$\mu$m respectively i.e. a fall of 10--15\% over a period of
$\sim$400~d. We found that the absolute flux values could be
influenced by the precise choice and placement of the aperture.
Suspecting that part of the dispersion in values arose from the
effects of aperture placement uncertainty on randomly orientated
images, we then aligned all the images to the same orientation and
repeated the aperture photometry.  The alignment procedure comprised
rotation and $x$, $y$ shifts, making use of the four or five most
compact isolated sources near the SN, per frame. This resulted in rms
values between 0\farcs02 and 0\farcs08 for the alignment
solutions. The 590~d, 684~d and 758~d data were aligned to the 994~d
data. The results of the photometry of the supernova region using
these aligned, unsubtracted images are given in Table \ref{tab3} and Figure
\ref{fig3}.  The photometric errors are slightly reduced relative to
those for the unaligned images, but temporal trends in the four
channels are much the same. \\

In our third approach, we aimed to achieve a more sensitive measure of
any time-dependent behaviour through the use of image matching and
subtraction techniques as implemented in the ISIS 2.2 image
subtraction package (cf. \citet{ala98}; \citet{ala00}). To provide a
better control of the image matching procedure, the original image
subtraction code was first modified to allow the user to input the
centre coordinates for the regions used for matching the images (see
also \citet{mat01}). We selected twelve $24\arcsec \times 24\arcsec$
regions centered on some of the brightest and most point-like sources
within 120\arcsec~ of the supernova, avoiding any artifacts present in
the backgrounds of the frames (the 2MASS star was not included due to
its proximity to the SN). We employed no background variability for
the backgrounds between the frames, and order 0, 1, and 2 variability
for the convolution kernel.  While the IRAC PSF is not expected to
vary with time, the shifting and rotating procedures applied to the
590~d, 684~d and 758~d data resulted in their PSFs being widened
relative to those of the 994~d images.  We therefore convolved the
994~d images to match each of the other epochs in turn.  We found that
the subtracted images showed a significantly lower level of residuals
at the location of bright sources as we increased the order of spatial
kernel variability from 0 to 1. This is because a spatially varying
kernel is able to correct for small imperfections in the image
alignment \citep{ala00}. However, only a modest improvement was found
going from order 1 to 2.  We therefore made use of the order~1
variability for all the images.  The 994~d images were then convolved
with a $17\arcsec \times 17\arcsec$ spatially varying kernel and their
intensities and backgrounds were matched with the aligned images.
Finally, the matched 994~d reference image was subtracted from each
aligned image to yield the difference in the SN flux between any given
epoch and the reference image epoch.  An example from the procedure is
given in Figure \ref{fig1} where we show the image subtraction of the 994~d
8.0~$\mu$m image from that at 590~d.  A bright point-like source is
clearly visible in the subtracted image and, to within the uncertainty
of the WCS co-ordinates, it lies at the supernova position. Comparison
of the source FWHM with those of isolated stars confirms that it is
indeed a point source.  We conclude that the subtracted-image point
source is entirely due to the supernova.  The accuracy of the
subtraction procedure is clear from the faintness of the residual
emission in the field.\\

\begin{figure}
\epsscale{1.0}
\plotone{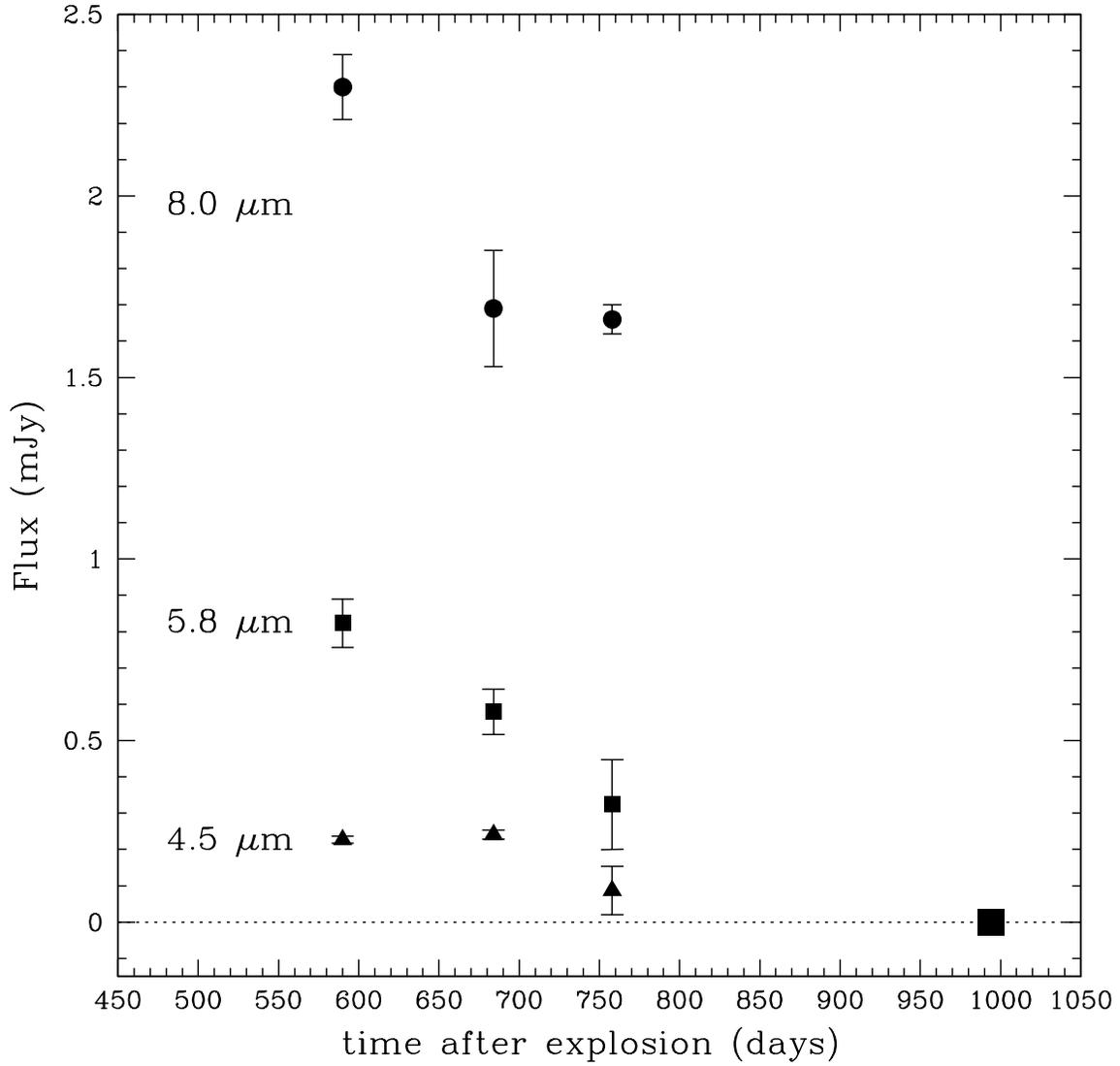}
\caption[]{Mid-IR photometry of SN~2002hh using
a 6\farcs5 diameter aperture centred at the supernova
position in subtracted images (see text). The square point
at zero flux represents the 994~d level for all four channels.  The
images at 994~d were subtracted from the corresponding channels at
the other epochs.
\label{fig4}
}
\end{figure}

Aperture photometry of the supernova in the subtracted images was then
carried out using the same aperture size and positioning as before.
The subtraction procedure achieved a good match between the background
levels of the two frames, yielding net background levels close to zero
in the subtracted frames.  Consequently, a bonus of using subtracted
images was that no background annuli were needed in the photometry
thus reducing possible contamination by imperfectly subtracted sources
lying within the annuli.  The uncertainties in the photometry include
the effects of photon statistics, aperture positioning error, and
imperfections in the image matching procedure. To investigate the
uncertainties introduced by image matching, we repeated the image
matching and subtraction procedure using another set of regions (with
some overlap with the first one) for fitting the convolution kernel
and background. We also compared the subtraction results for the MISC
and SINGS data on 758~d.  From this we conclude that systematic
uncertainties due to the image matching procedure dominate the errors
in the supernova photometry.  At 3.6~$\mu$m no significant change in
flux between any epoch was detected at the level of $\sim$0.2~mJy
(2~$\sigma$). This is in spite of a small apparent brightening seen in
the unsubtracted image fluxes with a photometry error of just
$\pm0.05$~mJy. We attribute the increased error in the 3.6~$\mu$m
subtracted image photometry to a very strong residual from the 2MASS
star produced during the subtraction process, possibly due to
saturation in the vicinity of the star.  However, in the other three
channels, as hoped, the uncertainty was reduced relative to the
unsubtracted image photometry. At 5.8~$\mu$m and 8.0~$\mu$m we see
error reduction of factors of $\sim\times3-\times8$.  The results are
shown in Table \ref{tab4} and Figure \ref{fig4}.  The 4.5, 5.8 and
8.0~$\mu$m all exhibited small but significant changes from epoch to
epoch.  Between the first and last epochs, declines of about 10\% are
seen relative to the total flux.  This confirms the suspected declines
in the unsubtracted images.  At 4.5~$\mu$m, there is little change
between 758~d and 994~d.  We suspect that, by this epoch, the
4.5~$\mu$m flux had faded below detectability.  However, at 5.8~$\mu$m
\& 8.0~$\mu$m there is no evidence to suggest that the observable
decline had ceased by the last epoch (994~d). The likelihood of a
subsequent decline is considered below in the context of the IR~echo
model.\\

The important conclusion that can be drawn from our demonstration of
temporal variation in the mid-IR fluxes is that at least a small
fraction of the total mid-IR flux from a region within $\sim$100~pc of
SN~2002hh must have been powered by the supernova.  In the following
section we discuss what fraction of the total flux in the aperture
might be reasonably attributed to the supernova, and what its cause
might be. \\

\section{Analysis and Discussion}
Interpretation of the mid-IR measurements of SN~2002hh depends
critically on how much of the flux is actually from the supernova
ejecta and/or nearby material, and how much is from causally
unconnected line-of-sight sources in NGC~6946.  In the following, we
consider three possible supernova-driven sources viz. newly-condensed
ejecta dust, an IR~echo from surrounding dust, and an ejecta/dusty CSM
interaction.  Evidence of mid-IR emission from dust associated with
SN~2002hh has been discussed by \citet{bar05} and \citet{mei05}.

\subsection{Mid-IR emission from newly-condensed ejecta dust}
In Figure \ref{fig5} (round points) we plot the dereddened (see below)
spectral energy distribution (SED) for the total flux in the aperture
(unsubtracted, aligned images) for epoch 590~d.  The SED is roughly
flat between 3.6~$\mu$m and 4.5~$\mu$m but then rises swiftly towards
longer wavelengths. This suggests a combination of two spectra. We
suggest that the shorter wavelength emission is due to a combination
of nebular emission from the ejecta plus residual stellar/nebular
background (recall that the aperture radius is equivalent to
$\sim$100~pc) while the longer wavelength component indicates
something much cooler.  The obvious candidate source is warm dust. We
have explored this by matching blackbodies to the 4.5, 5.8 and
8.0~$\mu$m fluxes.  In this procedure, the IRAC fluxes were first
dereddened according to the 2-component extinction of \citet{poz06},
extrapolated to the mid-IR using the \citet{car89} law. This yielded
scaling factors of $\times$1.191, $\times$1.133, $\times$1.089 and
$\times$1.053 for, respectively 3.6, 4.5, 5.8 and 8.0~$\mu$m. It might
be argued that such an extrapolation to this wavelength region is
unjustified.  \citet{dra03} gives a range of possible extinction laws
in the mid-IR.  In particular, the 8.0~$\mu$m extinction may be higher
than the Cardelli et al. law extrapolation suggests, due to the
9.7~$\mu$m silicate absorption. However, on the basis of Figure
\ref{fig4} in \citet{dra03} we estimate that other possible extinction
laws would change the mantissae of the mid-IR dereddening factors by
only around 10--15\%.  Good matches to the dereddened fluxes are
obtained with a temperature of 320~K at all epochs, and a blackbody
radius of $\sim9\times10^{16}$~cm at 590~d declining to
$\sim8\times10^{16}$~cm at 994~d.  The blackbody match for 590~d is
shown in Figure \ref{fig5}.  For this blackbody to reach a radius of
$9\times10^{16}$~cm requires a velocity of 18,000~km/s after the
supernova exploded.  The formation of dust at this velocity is highly
implausible.  Such high velocities are generally only seen in the
extreme outer zones of the H/He envelope. Furthermore, the extreme
wings of the brightest, earliest, post-plateau (epoch $\sim$150~d)
metal lines in SN~2002hh correspond to velocities of no more than
$\sim$4500~km/s \citep{poz06}.  (We ignore the strong, broad calcium
triplet P~Cygni feature present in the 44~d spectrum.  Given the
intrinsic strength of the Ca~triplet transition plus the fact that
this epoch is less than halfway through the plateau phase, the feature
is presumably caused by a small amount of pre-explosion calcium in the
progenitor atmosphere).  We therefore rule out ejecta dust, whether
heated by radioactivity or a reverse shock from a CSM impact, as the
source of the total flux in the 3\farcs3 radius aperture. A similar
argument against newly-formed dust was made by \citet{bar05} and
\citet{mei05}. \\

An additional argument against radioactively-heated ejecta dust being
the source of the total mid-IR flux in the aperture is provided by
energy considerations.  Integrating over the single temperature
blackbody matches, the total luminosity at 590~d is
$\sim6\times10^{40}$~erg/s.  SN~1987A produced a similar mass of
$^{56}$Ni to that seen in SN~2002hh \citep{poz06}.  (We assume that
the masses of other radioactive materials (e.g. $^{57}$Ni) were also
similar between the two events.)  Assuming a similar deposition
fraction in SN~2002hh, at 590~d radioactive decay would deposit only
$\sim3\times10^{39}$~erg/s in the ejecta \citep{li93}, falling to
about 2\% of this by 994~d. \\

While the above discussion rules out condensing dust as the source of
the total IR flux in the aperture, emission from newly-condensed
grains might be responsible for the temporally varying component.  In
Figure \ref{fig5} (square points) we also plot the SED for the
supernova obtained from the subtracted images for epochs 590, 684 and
758~d. The lower flux in this case means that the blackbody radius is
much reduced. Nevertheless, to obtain a good match we require a
blackbody velocity of $\sim$4800~km/s, with temperatures of 345~K,
315~K and 295~K for each epoch respectively.  Even this lower velocity
exceeds slightly that of the fastest metals seen in the post-plateau
SN~2002hh spectra (see above).  Moreover, the mass of dust required to
create a blackbody at 8$\mu$m lying at 4800~km/s is large.  A simple
calculation suggests that a mass of at least $\sim$0.005~M$_{\odot}$
would be needed (assuming a grain radius radius $\sim0.1~\mu$m,
material density $\rho=3$~g~cm$^{-3}$ and absorption/emission
efficiency $\propto \lambda^{-1}$).  It seems unlikely that so much
dust could form at such a high velocity. \\

As in the total IR~flux case, an additional argument against
radioactively-heated ejecta dust being the source of the declining
mid-IR flux in the aperture is provided by energy considerations.
Assuming a similar deposition of radioactive decay energy in SNe~1987A
\citep{li93} and 2002hh (see above), by 758~d the radioactive decay
deposition luminosity in SN~2002hh would be only 35\% of the blackbody
luminosity in the 0--10$\mu$m range.  Indeed, the luminosities of the
subtracted-image SN~2002hh 0--10$\mu$m integrated blackbody spectra
are $\times3-5$ greater than the coeval thermal emission from
condensing dust in SN~1987A \citep{woo93}.  Yet, as we have indicated,
the masses of radioactive material in the two SNe appear to be
similar.  In Figure \ref{fig6} we compare the subtracted-image
8.0~$\mu$m fluxes with the 8.4~$\mu$m light curve of SN~1987A, scaled
to the SN~2002hh distance.  The 8.4~$\mu$m emission from SN~1987A
during this phase is generally accepted as being mainly from
newly-condensed dust in the ejecta.  Between 550~d and 1000~d it can
be seen that the 8~$\mu$m energy emitted by SN~2002hh was about
$\times$6 greater.  We also recall that the subtracted-image fluxes
probably only represent lower limits for the amount by which the
SN~2002hh mid-IR flux finally declined.  The size of this decline is
estimated in 4.2.2 using an IR~echo model, and we plan to measure this
directly via later-epoch mid-IR images from {\it Spitzer}.  These
points argue against radioactively-heated ejecta dust as the main
source of the declining component. \\

Reverse-shock heating of ejecta dust is also unlikely since, as late
as 397~d, there was no evidence in the SN~2002hh optical spectra
\citep{poz06} of strong ejecta/CSM interaction as was seen, for
example, in the Type~IIn SN~1998S \citep{poz04}.  Moreover, in optical
spectra of SN~2002hh covering the period 650--1050~d \citep{cla05}
there is no sign of blueshifts in the H$\alpha$ line profiles such as
might have occurred if ejecta dust condensation had taken place.  We
conclude that IR emission from newly-formed ejecta dust is unlikely to
be the main cause of the declining component, whether heated by
radioactivity or a reverse shock from a CSM impact.\\

\subsection{Mid-IR emission from an IR~echo}
The total late-time mid-IR flux from the region of SN~2002hh could
have been due to thermal emission from nearby dust following heating
by supernova radiation emitted around the time of maximum light, i.e.,
an IR~echo.  \citet{poz06} have presented evidence from ground-based
observations during the first year that a dusty CSM existed around
SN~2002hh. Owing to the light travel time across the dust region, the
IR~echo emission seen at earth at a given time originates from a shell
bounded by ellipsoidal surfaces, with the axis coincident with the
line-of-sight. The thickness of this zone is fixed by the
characteristic width of the SN bolometric light curve, which is
dominated by UV-optical radiation. In addition, the peak light will
evaporate dust out to a certain distance.  \citet{dwe83,dwe85} has
shown that for a supernova with a peak UV-optical luminosity of
$1\times10^{10}~L_\odot$ and an exponential decline rate timescale of
25~d, the dust-free cavity radius is about $6\times10^{16}$~cm for
carbon-rich (graphite) grains (T$_{evap}=1900~K$) and
$3\times10^{17}$~cm for oxygen-rich (silicate) grains
(T$_{evap}=1500~K$).  This is for 0.1~$\mu$m radius particles. Scaling
to SN~2002hh using the early-time UV-optical light curve of
\citet{poz06} (see below), we infer a dust-free cavity of radius
$2.6\times10^{16}$~cm and $1.3\times10^{17}$~cm for graphite and
silicate grains, respectively.  While the UV-optical ellipsoidal shell
is still partially within the dust-free cavity, the IR light curve
should be characteristically flat \citep{dwe83,gra86,ger02}.  This is
due to the combined effects of the grain equilibrium temperatures, the
cloud geometry and the width and propagation of the ellipsoidal IR
emission region.  However, once the whole ellipsoidal shell has left
the cavity, the IR flux declines. \\

We have constructed a simple IR~echo model to test the possibility
that the total mid-IR flux within the 3\farcs3 radius aperture
originates in CSM dust heated by the peak SN luminosity.  (Molecular
cloud dust will be discussed later). In particular, we sought the
lowest CSM dust mass that could provide the observed flux.  The
IR~echo model follows those of \citet{bod80}, \citet{dwe83} and
\citet{gra86}. The model assumes a spherically-symmetric dust cloud
having a single grain size, with the actual value of the grain radius
as a free parameter.  However, we have also explored the effect of a
dust size distribution and this will be briefly described later.
Following \citet{poz06}, the input UV-optical luminosity is a
parameterised version of the peak and plateau parts of the SN~1999em
$UBVRI$ lightcurve \citep{elm03} corrected to the SN~1999em Cepheid
distance of 11.7~Mpc \citep{leo03} and then scaled to the 5.9~Mpc
distance of SN~2002hh.  To allow for radiation shortward of the
$U$-band, we assumed that the supernova spectrum could be roughly
described by a blackbody at 12,000~K during the peak, and 5,500~K
during the plateau. Therefore, we scaled the adopted light curve by
$\times$1.9 in the peak and $\times$1.05 in the plateau.  (For modest
changes ($\Delta T\sim1000$~K) in the adopted supernova peak
temperature the derived dust mass varies roughly inversely with the
temperature.)  At wavelengths longer than the $I$~band the fractions
of the total blackbody radiation are about 5\% at 12,000~K and 30\% at
5500~K. However, at these wavelengths the absorptivity of the dust
grains (size $\sim$0.1~$\mu$m) is likely to have fallen well below
unity.  Therefore, to simplify the IR~echo calculation, we ignored the
IR contribution to the total SN luminosity. We also ignored the
radioactive tail since this is (a) much fainter than the earlier
phases ($<$10\% of the peak flux) and (b) dominated by IR
radiation. Thus, the input UV-optical light curve used for the model
is: $47.0\times10^{41}$e$^{-t(d)/18.7}$~ergs/s to 24~d and
$11.1\times10^{41}$e$^{-t(d)/171}$~ergs/s for 24--118~d.  A grain
material density of 3~g~cm$^{-3}$ was assumed.  Given the typical
grain size ($\sim$0.1~$\mu$m) required to reproduce the SED and still
maintain a minimal dust mass it is possible that, as for the near-IR,
the absorptivity was below unity over at least part of the optical
range of the SN peak luminosity spectrum.  Therefore, for simplicity
we adopted a single grain absorptivity to the input UV/optical light
but allowed it to take values below unity. In calculating the mid-IR
emission from the dust, for wavelengths longer than $2\pi a$ grain
emissivities proportional to $\lambda^{-1}$ and $\lambda^{-2}$ were
considered.  Cloud density laws of $r^{-1}$ and $r^{-2}$ were used.
To estimate the mass of associated circumstellar gas one can use the
dust-to-gas mass ratio of the CSM.  However there is a wide range of
values found for evolved massive stars e.g. \citet{her05} quote a
range of 0.001--0.035 for oxygen-rich AGB stars. In view of this
uncertainty, CSM mass estimates will be expressed in terms of the
dust-to-gas mass ratio $r_{dg}$ scaled to the typical ISM value of
0.01. \\

To reproduce the very slow observed decline in the total mid-IR flux
(Fig.~3), it is necessary for the vertex of the outer echo ellipsoid
to remain within a dust-free cavity for a time $t/2$ where $t$ is the
time to the latest observation (994~d).  This corresponds to a cavity
size of $ct/2$=$1.3\times10^{18}$~cm. As indicated above, the maximum
likely size of an evaporated dust-free cavity would be only
$\sim1.3\times10^{17}$~cm.  Thus, the only way in which a CSM cavity
of the required size could occur would be via episodic mass loss.  We
adjusted the cavity radius to provide the best match to the light
curve.  This was obtained with an inner radius of
$1.25\times10^{18}$~cm. The outer radius was set at $\times$2 of this
value, although this is fairly uncritical.  For a given combination of
emissivity and density law, the grain size and emissivity were varied
until the model SED matched that of the observed 4.5--8.0~$\mu$m
fluxes.  The 3.6~$\mu$m flux showed a large excess with respect to the
model.  As suggested above, most of this emission is probably due to a
combination of nebular emission from the ejecta plus residual
stellar/nebular background.  The grain number density was then
adjusted to match the total observed fluxes. With a $\lambda^{-1}$
emissivity and a $r^{-2}$ (steady wind) density law, a fair match to
the data could only be achieved by increasing the dust mass to
0.47~M$_{\odot}$.  This corresponds to a cloud mass of
$47(0.01/r_{dg})~M_{\odot}$ where $r_{dg}$ is the dust-to-gas mass
ratio. The grain radius was 0.15~$\mu$m and absorptivity was 1. (We
note that the absorptivity would need to be reduced if the SN~2002hh
distance is greater than that adopted.)  This yielded a dust
temperature at the inner boundary of 345~K during the echo-plateau
phase, and a UV-optical absorption optical depth to the input SN
luminosity of $\tau_{abs}=0.42$.  Use of a shallower density gradient
or steeper emissivity law tended to increase the dust mass.  The model
matches to the 590 and 684~d SEDs and to the 5.8~$\mu$m and 8.0~$\mu$m
light curves are shown in Figures \ref{fig7} and \ref{fig8},
respectively.  Even if we invoke a CSM dust-to-gas ratio towards the
extreme high end of the observed range, the CSM mass would still be
too large to have arisen from a progenitor mass loss phase.  We
conclude that the {\it total} mid-IR emission within the 3\farcs3
radius aperture could not have arisen from an IR~echo within a
progenitor CSM. At this stage the possibility remains that most of the
flux might have arisen from an IR~echo from dust in a surrounding
molecular cloud.  This is addressed in subsection~4.4.\\

While we have concluded that the bulk of the mid-IR emission was not
caused by a CSM-IR~echo, it is still possible that it was responsible
for a small part of the emission, specifically the declining
component.  We therefore repeated our CSM-IR~echo analysis, but this
time using the supernova fluxes from the subtracted images (Table
\ref{tab4}). Naturally, a much lower dust mass is required to
reproduce the observations. As before, we endeavoured to find the
parameters which would minimise the mass of dust required. In
particular, we increased the dust-free cavity radius and hence the
plateau phase of the IR~echo light curve as far as was compatible with
the data.  An additional free parameter was introduced to take into
account the likelihood that the decline of the flux did not cease at
994~d. The zero-flux level of the data was therefore allowed to vary
to optimise the match to the model light curve shape.  Good matches to
the data were obtained with a grain radius of 0.07~$\mu$m and
UV/optical absorptivity of 0.3. However, the dust mass is relatively
insensitive to the choice of grain size/absorptivity parameter pair.
The UV-optical absorption optical depth to the input SN luminosity was
$\tau_{abs}=0.038$.  Thus, most of the extinction of the SN light must
have occurred beyond the CSM.  The outer and inner radii of the dust
distribution are $0.86\times10^{18}$~cm and $2.0\times10^{18}$~cm,
respectively.  However, the outer radius is relatively uncritical for
the match to the data. A lower value would reduce the dust mass, and
vice versa.  Other parameters were as for the total-flux model.  In
Figure \ref{fig9} we show the IR~echo model matches to the observed
SED at 590~d and 684~d.  In Figure \ref{fig10}, the IR-model light
curves are compared with the observed evolution at 5.8 and 8.0~$\mu$m.
The light curve matches indicate that, between 590~d and 994~d, the SN
flux declined by $\sim$0.75 of its 590~d value i.e. by 994~d we
estimate that the flux from the echo was about 0.3~mJy at 5.8~$\mu$m
and 0.9~mJy at 8.0~$\mu$m. As mentioned earlier, we plan to measure
the post-994~d evolution directly via later-epoch mid-IR images from
{\it Spitzer}.  We infer a dust mass of 0.036~M$_{\odot}$,
corresponding to a CSM mass of $3.6(0.01/r_{dg})~M_{\odot}$. This
gives a plausible CSM mass for a range of possible dust-to-gas ratios.
The large radius of the inner boundary of the dust required to
minimise the dust mass means that it is necessary to invoke episodic
mass loss.  For a wind velocity of 10~km/s the radius implies that the
mass loss phase declined significantly about 28,000 years before the
supernova explosion.  This timescale can be reduced, but at the cost
of increasing the dust mass.  The key point is that the declining
component {\it can} be reproduced using an IR~echo from a CSM of just
a few solar masses.  However, if the supernova was embedded in a
molecular cloud, an alternative scenario is that the declining flux
was due to an IR~echo from dense ISM dust in the supernova
vicinity. \\

The above modelling was based on the assumption of a single grain
size.  It may be argued that a more realistic approach is to use a
plausible grain size distribution. We have explored the effect of
introducing into the IR~echo model a grain size distribution
$dn\propto a^{-3.5}da$ with $0.005~\mu m<a<0.25~\mu m$
\citep{mat77}. A grain emission/absorption efficiency proportional to
$\lambda^{-1}$ was adopted when $\lambda>2\pi a$, and constant
otherwise.  We find very similar results to those obtained with the
single grain size assumption.  The derived dust masses were about
5--10\% lower than in the single-sized grain case. \\

\subsection{Mid-IR emission from ejecta/CSM interaction}
We have also considered the possibility that the total mid-IR
luminosity was due to the impact of the ejecta on a dusty CSM.  As
increasing amounts of ejecta gradually encountered the CSM dust, the
relatively constant mid-IR flux might be produced.  However, for this
to work, the dust-free cavity evaporated by the peak luminosity must
not be so large that the ejecta could not have reached the surviving
dust by the beginning of the observations.  As explained above, we
estimate a dust-free cavity of radius between $2.6\times10^{16}$~cm
and $1.3\times10^{17}$~cm. However, the lower limit is somewhat
irrelevant here, since to provide the total observed SED at 590~d
would require a blackbody of radius $9\times10^{16}$~cm (see
above). As already indicated, to reach a distance of
$9\times10^{16}$~cm in just 590~d would take a velocity of
$\sim$18,000~km/s.  For a dust-free cavity of radius
$1.3\times10^{17}$~cm it would require 26,000~km/s.  Inclusion of
light-travel time effects would increase these velocities by a further
5--10\%.  Only the very fastest, outermost material is likely to be
moving with such velocities.  It is unlikely that sufficient energy
could be transfered to the CSM dust by such tenuous material to
account for the 590~d luminosity.  Moreover, to explain the observed
mid-IR flux and SED the dust would have to be optically thick {\it in
the mid-IR region} or at least close to being so. For an optical depth
of unity at $\sim$8~$\mu$m, the 2-component extinction scenario
derived by \citet{poz06} extrapolated to the $V$-band using
\citet{car89} indicates an $A_V\sim90$.  Even if we adopt the flatter
$\lambda^{-1}$ dependence used in our IR~echo model, the $V$-band
extinction would be at least $A_V=15$.  Yet \citet{poz06} observe only
$A_V=5.0$.  Moreover, they find that K~I absorption suggests that at
least $A_V\sim3$ of this is probably interstellar extinction in the
host galaxy and Milky Way. We conclude that the total mid-IR emission
is unlikely to be due to ejecta impact onto CSM dust or, indeed, onto
ISM dust. \\

We have also considered the possibility that just the declining
component of the mid-IR emission was due to ejecta/CSM impact.  As
shown above, to reproduce the 590~d declining-component emission the
minimum source size corresponds to a blackbody expanding at 4800
km/s. This is equivalent to a radius of $2.4\times10^{16}$~cm which is
only just below the minimum grain survival radius of
$2.6\times10^{16}$~cm (see above).  Given that higher ejecta
velocities exist, could it be that an ejecta/CSM impact is a plausible
alternative to an IR~echo mechanism?  To address this question,
suppose that to find enough ejecta kinetic energy to power the mid-IR
flux we must include all velocities down to 10,000~km/s i.e.  we
consider ejecta/CSM impact heating of dust lying at roughly twice the
above blackbody radius.  Let us assume that the emission from the dust
is proportional to $(1-e^{-\tau})r_{dust}^2$ where $\tau$ is the
optical depth and $r_{dust}$ is the dust cloud inner radius.  If we
assume that the ``blackbody'' optical depth at 8~$\mu$m is at least 2
then for dust at 10,000~km/s to produce the observed 8~$\mu$m emission
we would require an optical depth of about 0.25.  Adopting a
conservative $\lambda^{-1}$ dependence of the grain
absorption/emission efficiency, this implies an $A_V\sim4.0$.  This is
still substantially larger than the extinction which \citet{poz06}
attribute to dust local to the supernova.  Moreover, as noted above
(subsection~4.1), in optical spectra of SN~2002hh covering the period
650--1050~d \citep{cla05} there is no sign of blueshifts in the
H$\alpha$ line profiles such as might have occurred if optically thick
dust were present.  Also, as noted earlier, as late as 397~d there was
no evidence in the SN~2002hh optical spectra \citep{poz06} of strong
ejecta/CSM interaction.  Another possible problem is that, given the
increasing amounts of ejecta moving into the dust, it might be
difficult to account for the apparently quite rapid decline
($\sim$0.42~mag per 100~d) in the mid-IR lightcurve (Fig.~10).  We
conclude that, while not ruled out completely, it seems unlikely that
ejecta/CSM impact can account for most of the declining mid-IR
emission. \\

In spite of the above conclusions, we know from the radio observations
of \citet{bes05} that some interaction between the ejecta and
circumstellar material was taking place as late as 899~d. In addition,
we have acquired near-IR evidence for a late-time CSM/ejecta impact.
We used the TIFKAM IR imager at the 2.4~m telescope of the MDM
Observatory, Kitt Peak, to obtain deep $JHK$ images of the SN~2002hh
field at 926~d. In all three bands we see a weak but clearly-detected
point source within 0\farcs15 of the supernova position.  After
dereddening, the IR fluxes are 0.18, 0.11 and 0.18~mJy at $J$, $H$
and $K$ respectively, with an uncertainty of $\pm$0.01~mJy in each
band.  We have searched for this source in pre-explosion
images. Unfortunately, there are no $JHK$ images of equivalent depth
and resolution.  However a deep $i_{\rm CCD}$-band image taken under
fair seeing conditions (S.~Smartt, private communication) shows no
point source at the supernova location, with a 5~$\sigma$ limit of
22.3.  This translates to a dereddened limit of 0.015~mJy.  We
suggest, therefore, that the $JH$ emission was produced by the
supernova (the origin of the $K$ flux is discussed later).  These
fluxes are over $\times$100 greater than the $JH$ flux from SN~1987A
(scaled to 5.9~Mpc) at the same epoch. Indeed the $JH$ point source
luminosity exceeds the total likely radioactive deposition
\citep{li93} by a factor of $\times$10. We therefore eliminate
radioactively-driven ejecta emission as the source of the $JH$
luminosity.  The dereddened $J-H$ is about zero, suggesting a
characteristic temperature of around 10,000~K. This rules out an
IR~echo source since the dust temperature is too high.  We therefore
suggest that the $JH$ flux is due to ejecta/CSM interaction.  Indeed,
a blackbody at this temperature would only need a velocity of $\sim$5
km/s to produce the observed $JH$ flux.  For the excess $K$~band
emission, we think a supernova origin is less likely.  Extrapolation
of the 18,000~km/s blackbody (cf. subsection 4.1) to the $K$~band,
with a slightly higher temperature of $\sim$400~K, shows that much of
the $K$~excess could simply be due to the Wien tail of the cool mid-IR
source.\\

\subsection{On the nature of the mid-IR source at the SN~2002hh location}
To explore further the origin of the bulk of the mid-IR flux, we have
compared the {\it Spitzer} 590~d 8~$\mu$m image with pre-explosion ISO
images of the same field \citep{rou01}.  In Figure \ref{fig11} we show the
8~$\mu$m (bandpass = 6.4--9.3~$\mu$m) {\it Spitzer} image placed
between the 7~$\mu$m (bandpass = 5.0--8.5~$\mu$m) and 15~$\mu$m
(bandpass = 12.0--18.0~$\mu$m) ISO images.  We have rebinned and
smoothed the 8~$\mu$m image to match the $\sim$6\arcsec~resolution of
the ISO images.  The contrast levels were set so that the galaxy
structure features are of similar contrast in each image. Further
details are given in the figure caption. As we go from short to long
wavelengths, the 2MASS star steadily fades as expected for a hot
object.  It dominates the supernova field at 7~$\mu$m but is less
prominent at 8~$\mu$m. By 15~$\mu$m the star is actually slightly
fainter than a source lying about 6\arcsec~ (2 pixels) east and
3\arcsec~ (1 pixel) south of it i.e. the images appear to be
consistent with an almost unvarying cool source lying to the east and
south of the 2MASS star, close to the supernova location.
Extrapolation from the 8~$\mu$m image to 7~$\mu$m assuming a
Rayleigh-Jeans slope for the 2MASS star spectrum and a 320~K blackbody
for the cool source (subsection~4.1) yields a total flux for the 2MASS
star + cool source which can be directly compared with that measured
in the 7~$\mu$m pre-explosion ISO image. Allowing for the
uncertainties in this extrapolation, we deduce that less than half of
the flux from the cool 8~$\mu$m source can be due to the supernova
explosion.  This suggests that {\it most} of the mid-IR flux from the
supernova direction was not actually due to the supernova i.e. most of
the flux was not due to an IR~echo from the CSM, supporting the
conclusions from our IR~echo modelling above, nor was it due to an
IR~echo from a surrounding dusty molecular cloud.  The {\it Spitzer}
images show that the region around the supernova is rich in cool,
mid-IR emitting structures (cf. Fig.~1), and we suggest that such a
source lying close to the line-of-sight is responsible for most of the
observed mid-IR flux. We propose that a luminous but highly-obscured
object, such as a star-formation region and/or molecular cloud could
be responsible.  If the supernova lies behind or is embedded in this
cloud it could explain the high extinction. In particular, it may
account for the anomalous extinction component identified by
\citet{poz06}. In addition, as suggested above (subsection 4.2), such
a cloud may also have produced the declining component of the mid-IR
flux via an IR~echo.  \\

Comparison of the {\it Spitzer} 3.6~$\mu$m flux with the earlier
$L'$~band evolution \citep{poz06} supports our contention that most of
the flux in the unsubtracted images was not due to the
supernova. Simple extrapolation of the $L'$~band light curve to 590~d
predicts a flux of 0.19~mJy. This is nicely consistent with the
0.2~mJy upper limit for the 3.6~$\mu$m subtracted images (Table
\ref{tab4}), but is less than a tenth of the 2.5~mJy in the
unsubtracted images.\\

\subsection{Comparison with the work of the SEEDS collaboration} 
The SEEDS collaboration \citep{bar05} have also presented flux
estimates for the SN~2002hh region in the SINGS images (see
Section~2), and we now compare their results with those presented
here.  They used a PSF fitting procedure to determine the fluxes. For
both SINGS epochs, at 4.5 and 5.8~$\mu$m, their fluxes are typically
$\sim$65\% of the values we obtained from the unsubtracted images.  At
8~$\mu$m their fluxes are, respectively, about 80\% and 60\% of our
values at 590~d and 758~d.  Given the complexity of the field and the
different photometry techniques used, these differences are not
surprising.  Between 590~d and 758~d we find that the 5.8~$\mu$m flux
fell by about $0.14\pm0.07$~mJy, compared with a $0.63\pm0.61$~mJy
decline reported by \citet{bar05}. While there appears to be a
disagreement, the difference is actually only at the level of
0.8~$\sigma$ when the large error in the latter value is taken into
account. Similarly, at 8.0~$\mu$m we find the flux fell by
$0.64\pm0.10$~mJy compared with $4.2\pm2.1$~mJy for \citet{bar05}.
Here, the disagreement seems even greater than at 5.8~$\mu$m, but
again the large SEEDS error means that the significance is only
1.7~$\sigma$.  Thus, there is actually fair consistency between the
two sets of results.  However, the present work indicates that,
relative to the {\it total} flux, the 8.0~$\mu$m 590--758~d decline
was only $\sim$4\% (scaled to the Barlow et al. total fluxes) and not
25\% as adopted by \citet{bar05}. \\

\citet{bar05} modelled the mid-IR flux using a dusty CSM heated by a
``central source''.  While their measured fluxes are comparable to the
values we obtain from the unsubtracted images, they infer a dust mass
of 0.10--0.15~M$_{\odot}$ i.e. a factor of $\sim$4 less than our
0.47~M$_{\odot}$ dust mass estimate based on the total fluxes.  A
possible reason for the large discrepancy between our respective
estimates is their apparent non-inclusion of light-travel time effects
in their models.  One consequence of light-travel time delays is that,
at any given time the bulk of the IR flux we receive may come from
only a relatively small fraction of the CSM, bounded by the echo
ellipsoids corresponding to the characteristic width of the early-time
light curve.  We therefore suggest that, assuming a spherically
symmetric CSM, Barlow et al. have underestimated the CSM mass required
to reproduce the total observed mid-IR flux via this mechanism.  In
addition, we note that \citet{bar05}'s inner CSM radius of only
$2.1\times10^{17}$~cm ($\sim$80~light-days) in their best-fitting
model is unlikely to reproduce the near-constant flux to 994~d.
Barlow et al. also present an 11.2~$\mu$m image of the supernova field
obtained at the 8m Gemini North telescope at 698~d. The spatial
resolution is 0\farcs3 (8.6~pc at NGC~6946) - considerably higher than
that of {\it Spitzer}.  A bright unresolved point source is
apparent at the supernova location, which they identify with the
supernova.  However, as argued above, we believe that the bulk of the
mid-IR flux in this wavelength region is not from the supernova, but
rather is due to a luminous, obscured star-formation region or
molecular cloud.  It is our view that most of the flux assumed by
\citet{bar05} to have been driven by the supernova was, in fact, from
a separate, causally-unconnected source.

\section{Summary} 
We have presented late-time mid-IR observations of the normal, but
highly-reddened Type~IIP SN~2002hh. This is only the second-ever
core-collapse supernova to be observed at such late epochs in the
mid-IR, and is the first Type IIP to be studied in this way.  The
4.5--8.0~$\mu$m flux from within 3\farcs3 ($\sim$100~pc) of the
supernova reveals bright, cool emission with a characteristic
temperature of about 320~K.  A flux decline of about 10\% was seen
between 590~d and 994~d.  We rule out condensing dust in the ejecta,
an ejecta/CSM impact or a CSM-IR~echo as the cause of most of the
observed mid-IR flux.  Comparison of the {\it Spitzer} data with
pre-explosion ISO images also tends to rule out an IR~echo from ISM
dust. This comparison suggests, rather, that most of the mid-IR flux
from the supernova direction is actually due to a luminous, cool
source such as a heavily-obscured star formation region or molecular
cloud lying along the line-of-sight.  This is supported by
extrapolation of the $L'$~band light curve.  The supernova may
actually lie behind or be embedded in the obscured source which would
account for the high extinction observed. We disagree with
\citet{bar05} that most of the mid-IR flux is due to the supernova and
with their suggestion that a large fraction of the extinction to
SN~2002hh appears to be due to CS dust.  \\

We assume that the small declining component of the mid-IR flux was
powered by SN~2002hh.  As possible sources we considered (a)
newly-condensed dust in the ejecta, (b) an IR~echo from a dusty CSM,
and (c) ejecta/dusty CSM impact.  Of these, an IR~echo provides the
most plausible explanation for most of the declining component. If the
supernova was embedded in the molecular cloud then it is possible that
the IR~echo was from dense ISM dust rather than from the progenitor
CSM.  Emission from condensing ejecta dust contributes, at most, a
small fraction of the declining flux.  This appears to be consistent
with the absence of late-time blueshifts in the H$\alpha$ line
profiles \citet{cla05}.  However, it may be that a substantial mass of
dust formed in the ejecta but is largely concealed in optically thick
regions.  A similar point has been made by \citet{bar05}.  Emission
from CSM dust heated in an ejecta/CSM impact is not completely ruled
out, but given the difficulties in reconciling this with the observed
optical extinction, the \citet{cla05} result, and, perhaps, the rapid
decline of the mid-IR flux, it seems that this mechanism also can only
provide a small part of the declining mid-IR flux. \\

We conclude that SN~2002hh produced an IR~echo from dust, either in
the progenitor CSM or in a surrounding molecular cloud, and that this
was responsible for most of the declining component of the mid-IR
flux. For the CSM scenario, we deduce that the mid-IR flux could have
been produced by a dust mass of as little as 0.036~M$_{\odot}$ lying
between $0.86\times10^{18}$~cm and $2.0\times10^{18}$~cm from the SN.
The corresponding CSM mass is $3.6(0.01/r_{dg})~M_{\odot}$.  This
gives a plausible CSM mass for a range of possible dust-to-gas ratios.
To account for the large inner boundary radius it is necessary to
invoke episodic mass loss.  For a wind velocity of 10~km/s, this
radius implies that the mass loss rate declined sharply around
28,000~years before the explosion.  \citet{poz06} used near-IR
observations of SN~2002hh during the first year to deduce a dusty CSM
mass of just 0.3~M$_{\odot}$, with an inner boundary at
$1.3\times10^{17}$~cm.  However, this does not necessarily conflict
with our mid-IR result as the \citet{poz06} study was sensitive only
to hot dust (T$\sim$1000~K) lying close to the supernova.  Such dust
could have formed in a much reduced mass-loss phase which took place
after the main mass-loss phase ended at 28,000 years ago. The small
amount of hot dust detected by \citet{poz06} would have been
undetectable by our {\it Spitzer} study. In addition, \citet{poz06}
would have been unable to detect the large CSM mass reported here as
it was too cold.  This scenario is consistent with the weak CSM
interaction inferred at early times by \citet{poo02} and
\citet{sto02}, and the low mass loss rate deduced from the early-time
radio observations by \citet{che06}. This ongoing ejecta/CSM collision
could also account for the very late-time $JH$-band source at the
supernova position.  The pre-explosion $i'$~band image
(subsection~4.3) allows an upper limit of 20M$_{\odot}$ to be placed
on the progenitor mass (S.~Smartt, private communication).  A star of
mass just below this limit could plausibly lose 3--10M$_{\odot}$
during its red supergiant phase, consistent with the
$3.6(0.01/r_{dg})~M_{\odot}$ CSM deduced from the IR~echo analysis.
For the molecular cloud IR~echo case, it is likely that a similar mass
would have been involved in producing the IR radiation, but this could
be just a small fraction of the total mass in the cloud.  Our IR~echo
analysis was based on the assumption of a spherically symmetric dust
cloud and we have shown that this was sufficient to account for the
mid-IR fluxes and evolution.  Consequently, asymmetric dust
distributions \citep{emm88} were not considered on this occasion.  \\

This work demonstrates for the first time that an IR~echo has occurred
in an apparently normal Type~IIP supernova, the commonest of all
supernovae.  We find no evidence for or against the proposition that
ejecta dust formation in such supernovae produces the 0.1-1
M$_{\odot}$ of grains required if such events are to be established as
a major contributor to the dust content of the universe.  However,
substantial masses of newly-condensed grains may lie concealed in
optically thick regions.  Observations carried out at later epochs and
longer wavelengths should allow us to decide if this is the case.  If
we opt for a CSM-IR~echo, the $\sim$0.04~M$_{\odot}$ of dust in the
progenitor wind is considerably larger than the
$\sim10^{-3}$~M$_{\odot}$ of directly-observed ejecta dust which has
been deduced from studies of SNe~1987A, 1998S and 1999em. Thus, while
we cannot yet decide if Type~IIP supernovae form large masses of
grains, it does appear that important contributions to the dust
content of the present universe could be made by their progenitor
winds.



\acknowledgments

We thank Christophe Alard for helpful discussions, Steve Smartt for
his progenitor mass limit estimate, and Mike Barlow for a computer
readable version of the 11.2~$\mu$m image of the SN~2002hh field.  CLG
was supported in part by PPARC Grant PPA/G/S /2003/00040.  RK was
supported in part by EU RTN Grant HPRN-CT-2002-00303.  SM acknowledges
financial support from the European Science Foundation.  MP is
supported by PPARC Grant PPA/G/S/2001/00512.  JCW is supported in part
by NSF Grant AST-0406740. This work is based on observations made with
the Spitzer Space Telescope, which is operated by the Jet Propulsion
Laboratory, California Institute of Technology under a contract with
NASA. Support for this work was provided by NASA through an award
issued by JPL/Caltech.



Facilities: \facility{Spitzer Space Telescope,~} \facility{SSC Leopard
Archive Tool,~} \facility{ NASA/IPAC Extragalactic Database ({\it NED})}




\clearpage





\clearpage

\input{tab1.tex}

\clearpage

\input{tab2.tex}

\clearpage

\input{tab3.tex}

\clearpage

\input{tab4.tex}

\clearpage



\begin{figure}
\epsscale{0.96}
\plotone{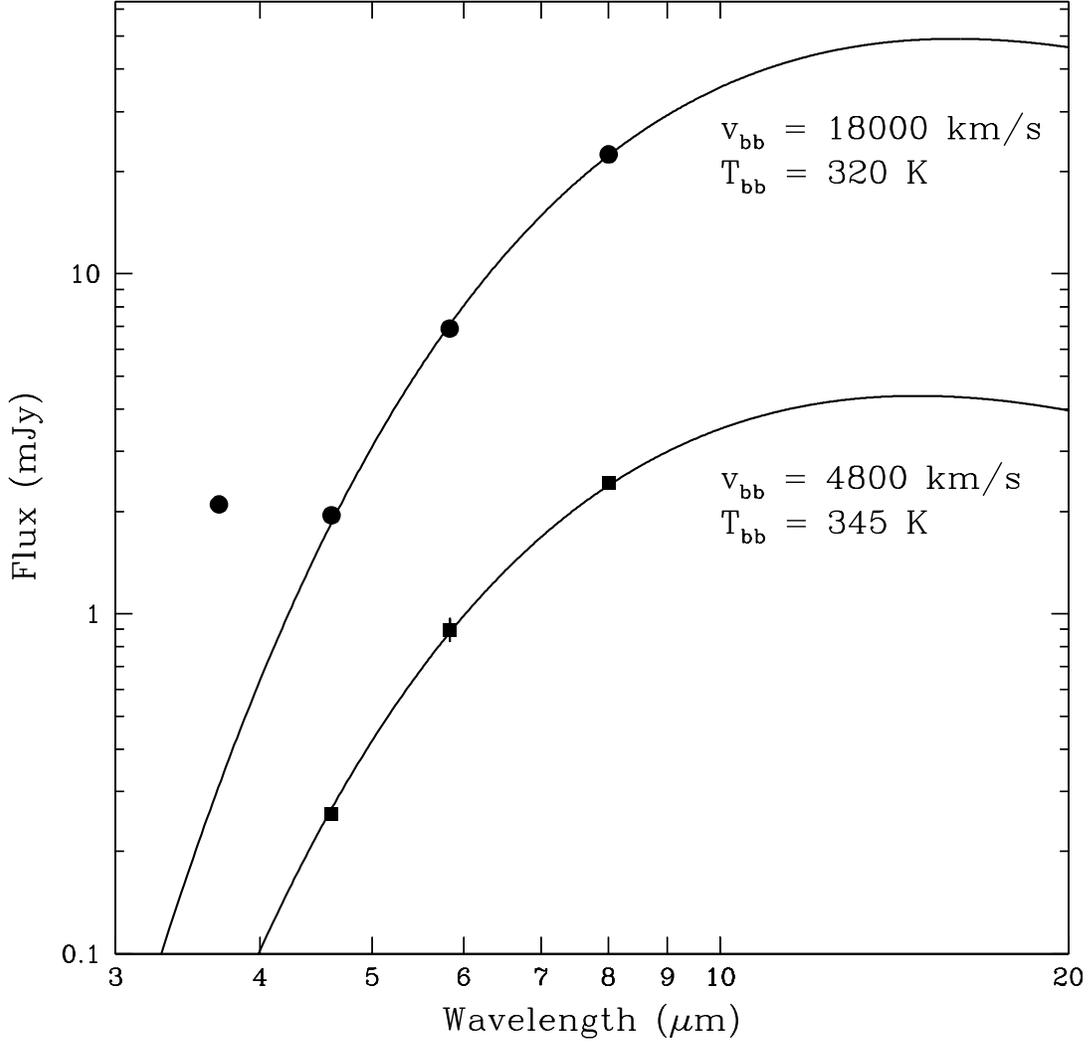}
\caption[]{IRAC photometry of the SN~2002hh region using a 6\farcs5
diameter aperture centred at the supernova position. The round points
show the fluxes obtained in the aligned, unsubtracted images at 590~d.
The square points show the fluxes obtained in the subtracted images
where the 994~d images have been subtracted from the 590~d images.
Each point has been dereddened according to the 2-component extinction
of \citet{poz06}, extrapolated to the mid-IR using the \citet{car89}
law. The individual points are located at the effective wavelengths of
the IRAC filters for a 325 K blackbody.  Also shown are blackbody
spectra adjusted to match the 4.5, 5.8 and 8~$\mu$m points for both
unsubtracted and subtracted cases.
\label{fig5}
}
\end{figure}

\begin{figure}
\epsscale{1.0}
\plotone{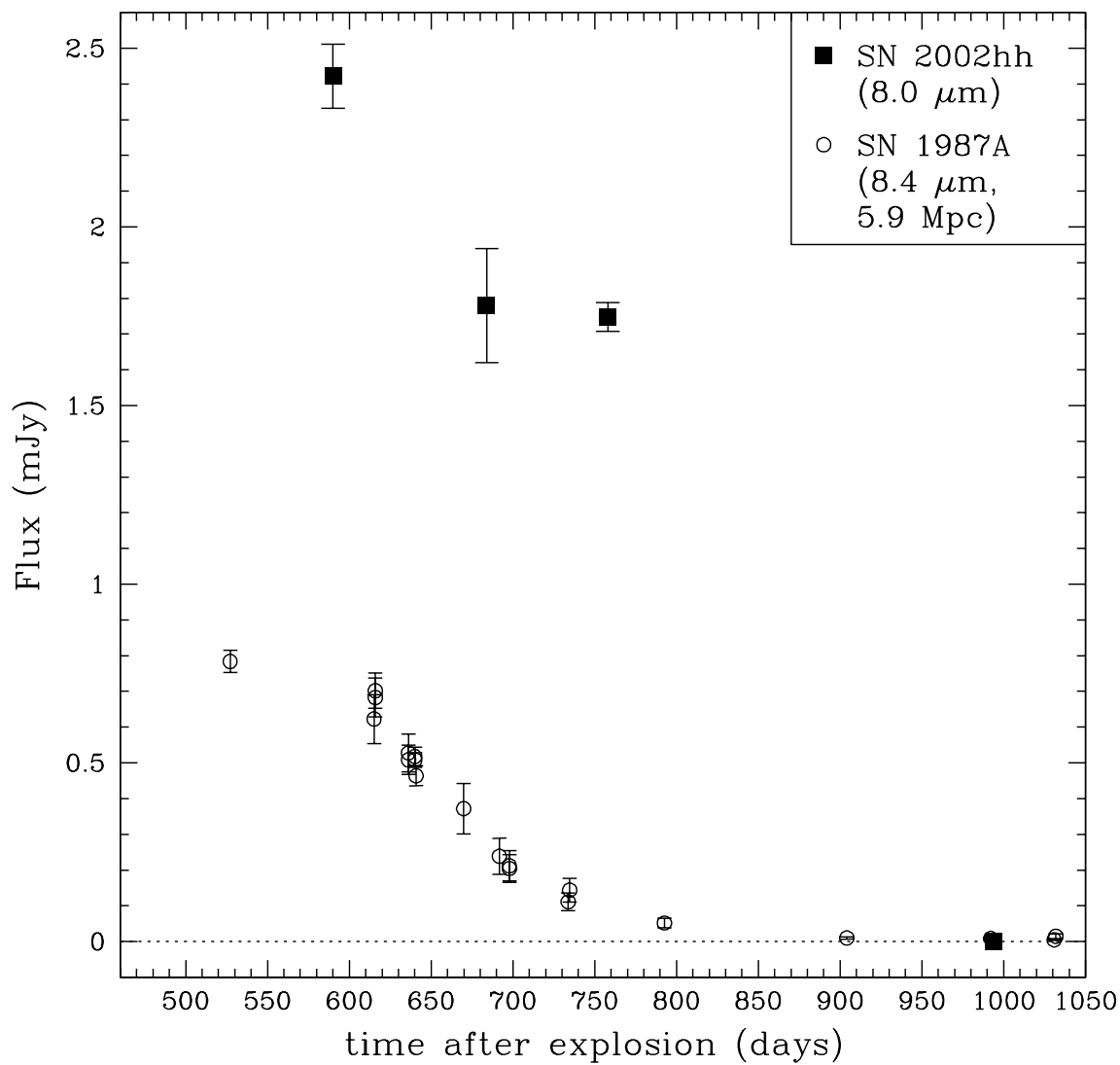}
\caption[]{8.0~$\mu$m photometry (square, filled points) of SN~2002hh
  using a 6\farcs5 diameter aperture centred at the supernova position
  in subtracted images. Each point has been dereddened according to
  the 2-component extinction of \citet{poz06}, extrapolated to
  the mid-IR using the \citet{car89} law.  The square point
  at zero flux represents the 994~d level. Also shown (round, open
  points) is the dereddened 8.4~$\mu$m light curve of SN~1987A,
  shifted to the SN~2002hh distance of 5.9~Mpc.
\label{fig6}
}
\end{figure}

\begin{figure}
\epsscale{1.0}
\plotone{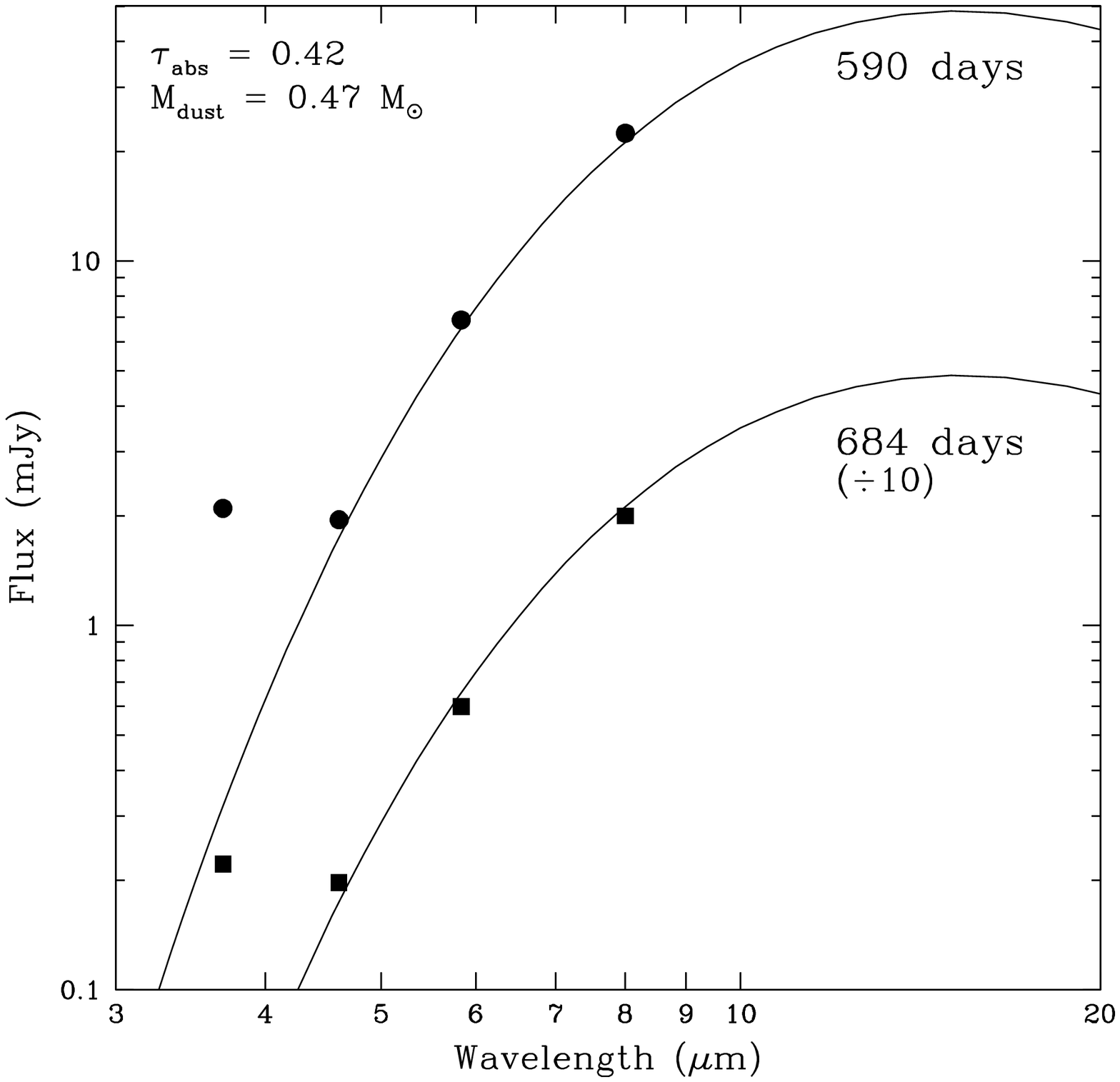}
\caption[]{Dereddened SED of the SN~2002hh region (from unsubtracted
images) at 590~d (round points) and 684~d (square points).  Each point
has been dereddened according to the 2-component extinction of
\citet{poz06}, extrapolated to the mid-IR using the \citet{car89}
law. The individual points are located at the effective wavelengths of
the IRAC filters for a 325 K blackbody.  The 684~d fluxes are shown
divided by 10 for clarity.  The curves are synthetic spectra produced
by an IR~echo model with a dust mass of 0.47~M$_{\odot}$ (i.e. a CSM
mass of $47(0.01/r_{dg})~M_{\odot}$).
\label{fig7}
}
\end{figure}

\begin{figure}
\epsscale{1.0}
\plotone{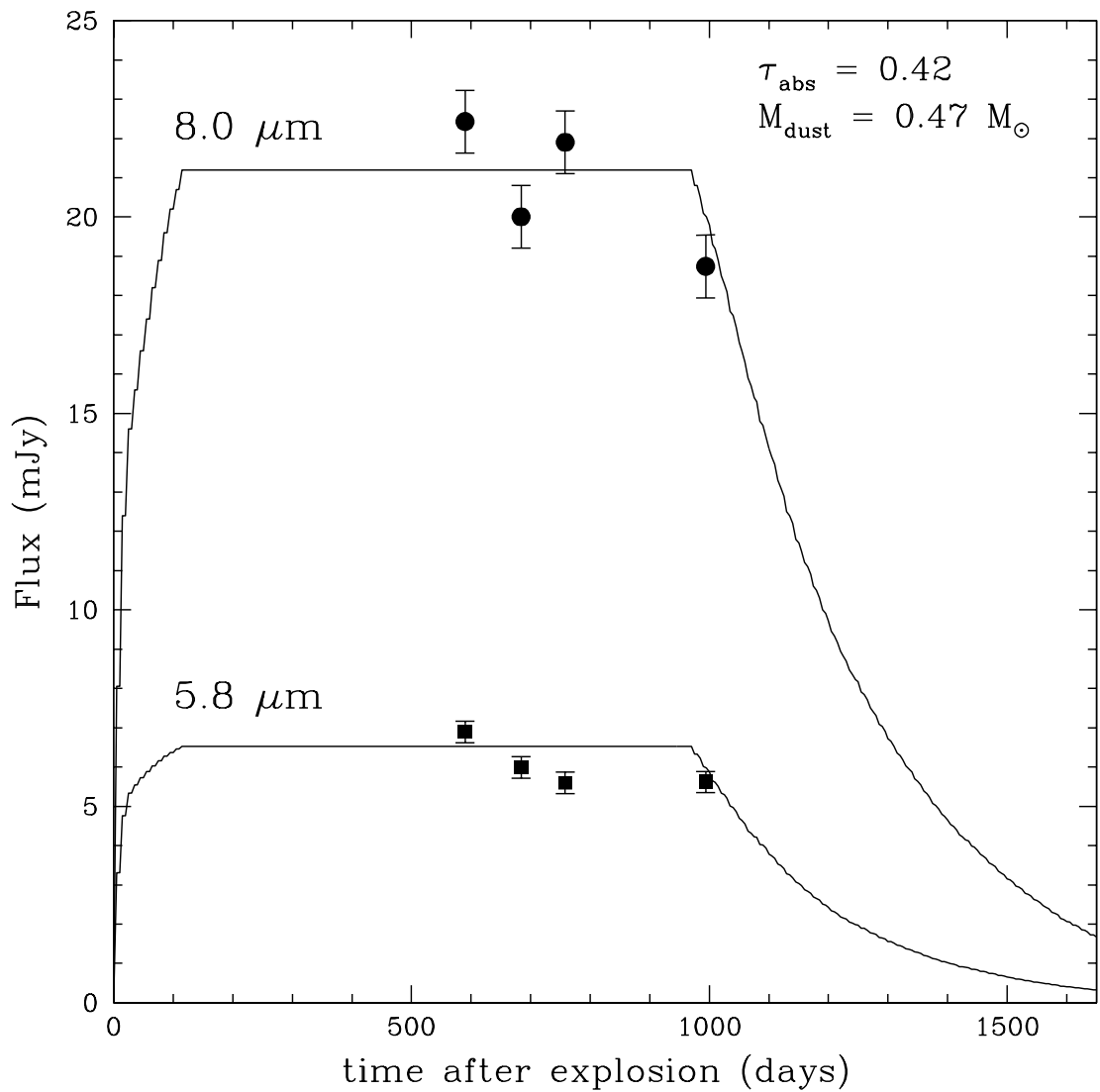}
\caption[]{ Dereddened 5.8~$\mu$m and 8.0~$\mu$m photometry of the
SN~2002hh region (from unsubtracted images) compared with synthetic
light curves produced by an IR~echo model with a dust mass of
0.47~M$_{\odot}$ (i.e. a CSM mass of $47(0.01/r_{dg})~M_{\odot}$). The
dust temperature at the inner boundary was 345~K during the IR~echo
plateau phase.
\label{fig8}
}
\end{figure}

\begin{figure}
\epsscale{0.925}
\plotone{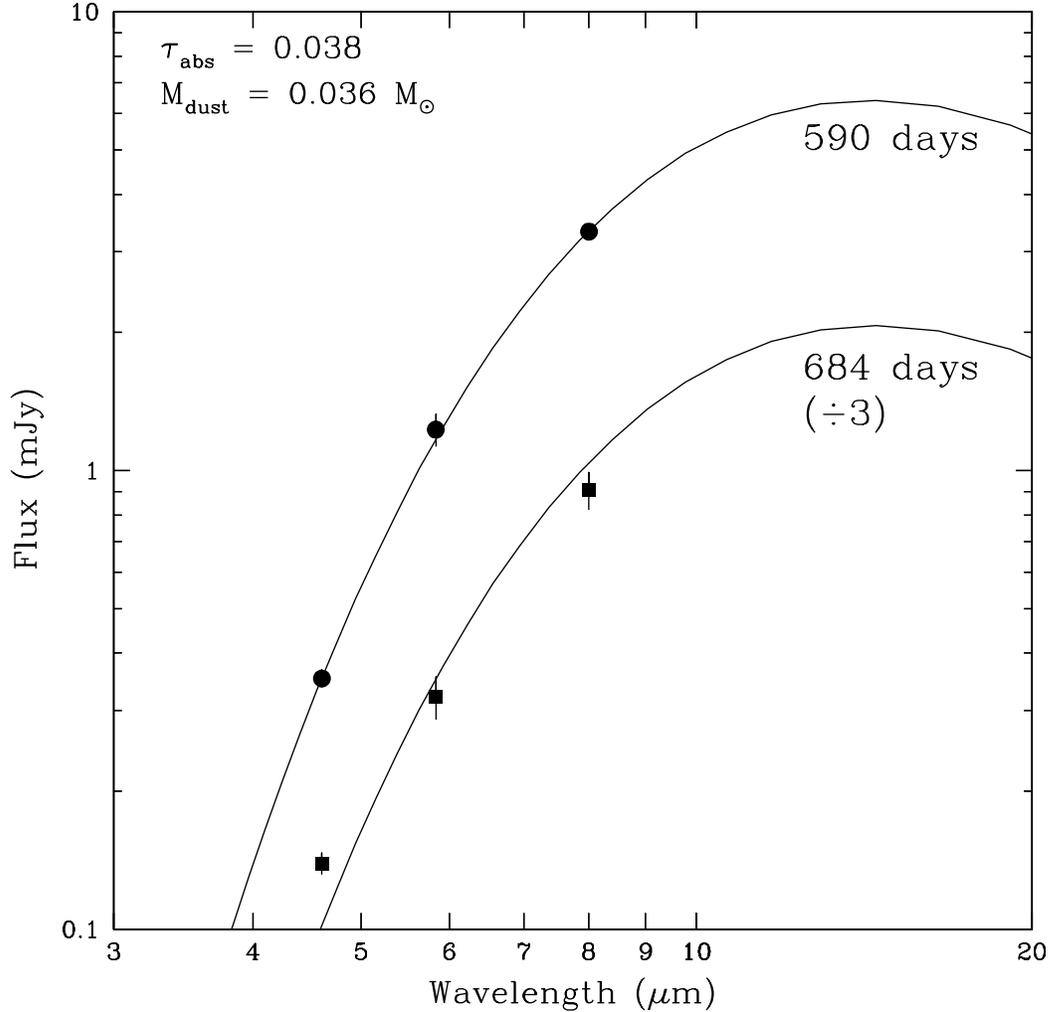}
\caption[]{Dereddened SED of SN~2002hh (from subtracted images)
at 590~d (round points) and 684~d (square points).  Each point has
been dereddened according to the 2-component extinction of
\citet{poz06}, extrapolated to the mid-IR using the \citet{car89}
law. The individual points are located at the effective wavelengths of
the IRAC filters for a 325~K blackbody.  The zero-flux level of the
data has been adjusted to optimise the match to the model light curve
shape (Figure~10).  The 684~d fluxes are shown divided by 3 for
clarity.  The curves are synthetic spectra produced by an IR~echo
model with a dust mass of 0.036~M$_{\odot}$ (i.e. a CSM mass of
$3.6(0.01/r_{dg})~M_{\odot}$). The dust temperature at the outer
ellipsoid vertex declined from 375~K the end of the IR~echo plateau
phase, to 315~K by 1000~d.
\label{fig9}
}
\end{figure}

\begin{figure}
\epsscale{1.0}
\plotone{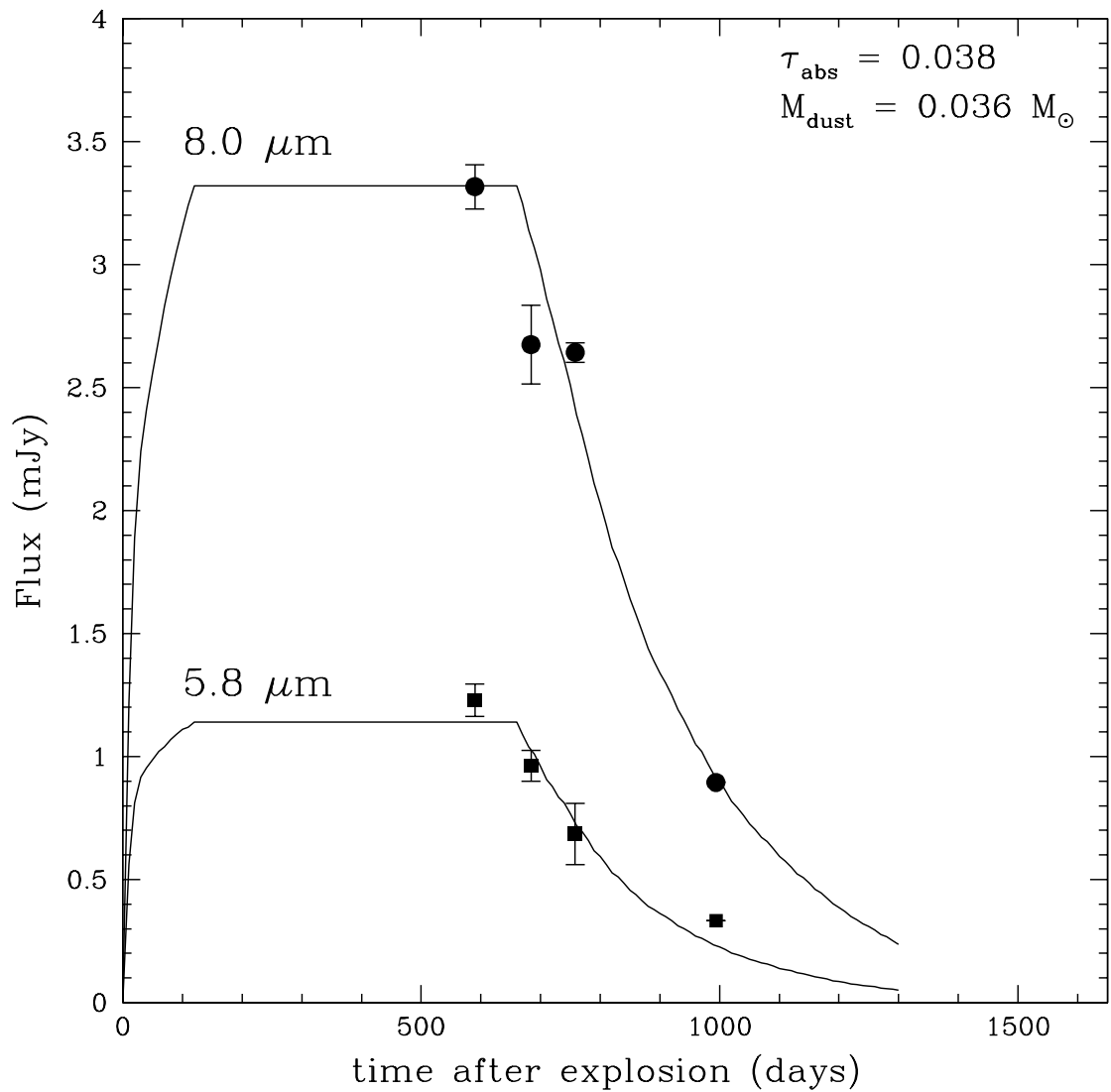}
\caption[]{Dereddened 5.8~$\mu$m (square points) and 8~$\mu$m (round
points) photometry of SN~2002hh (from subtracted images), compared
with synthetic light curves produced by an IR~echo model with a dust
mass of 0.036~M$_{\odot}$ (i.e. a CSM mass of
$3.6(0.01/r_{dg})~M_{\odot}$).  The zero-flux level of the data has
been adjusted to optimise the match to the model light curve shape. 
\label{fig10}
}
\end{figure}

\begin{figure}
\epsscale{1.0}
\plotone{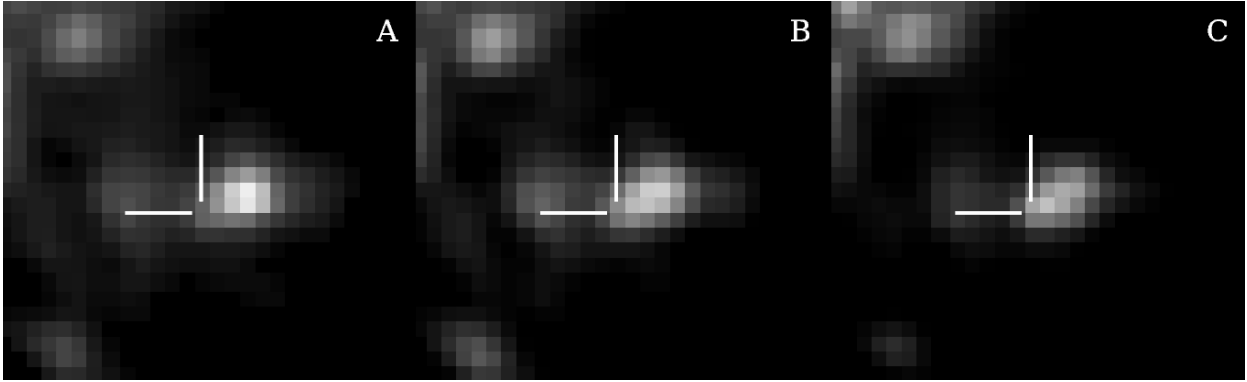}
\caption[]{Comparison of the 590~d 8~$\mu$m {\it Spitzer} image of the
SN~2002hh field {\it (B)} with pre-explosion ISO images (from
\citet{rou01}) of the same field at, respectively, 7~$\mu$m {\it (A)}
and 15~$\mu$m {\it (C)}.  The image FOV of each image is $82\arcsec
\times 76\arcsec$. North is up and East to the left.  The pixel size
is 3\arcsec~.  The SN location is marked with ticks in each image.
The 2MASS star lies about 9\arcsec~ from the supernova location at
P.A. 298~deg.  As we go from short to long wavelengths, the 2MASS star
steadily fades as expected for a hot object.  By 15~$\mu$m the star is
slightly fainter than a source lying about 6\arcsec~ east and
3\arcsec~ south of it.  i.e. the images appear to be consistent with
an almost unvarying cool source lying to the east and south of the
2MASS star, close to the supernova location.\\ To prepare this figure,
the ISO 7~$\mu$m image was first rotated to align the pixel and NE
axes.  The {\it Spitzer} image was then aligned and rebinned to match the
7~$\mu$m image. The aligned {\it Spitzer} image was then smoothed with a
Gaussian kernel with sigma = 1 pixel, to match the ISO resolution.
Finally, the 15~$\mu$m ISO frame was rotated and shifted to match the
aligned {\it Spitzer} frame using the positions of 3 sources around the SN,
including the blend between the 2MASS star and the nearby cool
source.
\label{fig11}
}
\end{figure}


\end{document}

%% file: tab1.tex
\begin{table}
\tablenum{1}
\begin{center}
\caption{Log of mid-IR (IRAC) observations of SN~2002hh with the Spitzer Space Telescope.}
\begin{tabular}{ccccl}
\tableline\tableline
JD  & Date          & {Epoch\tablenotemark{a}} & {t$_{int}$\tablenotemark{b}} & Program \\ 
(2453000+)   &  (UT)         &  (d) & (s)     &  \\ \hline
167.3        & 2004 Jun 10.8 & 590 & 214 & SINGS  \\
260.8        & 2004 Sep 12.3 & 684 & 536 & MISC \\
335.2        & 2004 Nov 25.7 & 758 & 214 & SINGS \\
335.3        & 2004 Nov 25.8 & 758 & 536 & MISC \\ 
571.7        & 2005 Jul 20.2 & 994 & 536 & MISC \\ 
\tableline
\end{tabular}
\tablenotetext{a}{Days after explosion (JD 2452577).}
\tablenotetext{b}{Integration time per filter.}
\end{center}
\label{tab1}
\end{table}

%% file: tab2.tex
\begin{table}
\tablenum{2}
\begin{center}
\caption{Mid-IR fluxes of the SN~2002hh region - unaligned, unsubtracted images.}
\begin{tabular}{ccccc}
\tableline\tableline
Epoch& \multicolumn{4}{c}{Flux (mJy)} \\ 
(d) & 3.6~$\mu$m & 4.5~$\mu$m & 5.8~$\mu$m& 8.0~$\mu$m \\ \hline
590 & 1.78(0.06)\tablenotemark{a} & 1.78(0.04) & 6.51(0.28) & 21.5(0.9)\\
684 & 1.90(0.06) & 1.81(0.04) & 5.57(0.28) & 19.2(0.9)\\
758 & 2.13(0.06) & 1.60(0.04) & 5.24(0.28) & 21.0(0.9)\\ 
994 & 2.00(0.06) & 1.57(0.04) & 5.31(0.28) & 18.0(0.9)\\
\tableline
\end{tabular}
\tablenotetext{a}{Errors are shown in brackets}
\end{center}
\label{tab2}
\end{table}

%% file: tab3.tex
\begin{table}
\tablenum{3}
\begin{center}
\caption{Mid-IR fluxes of the SN~2002hh region - aligned, unsubtracted images.}
\begin{tabular}{ccccc}
\tableline\tableline
Epoch& \multicolumn{4}{c}{Flux (mJy)} \\ 
(d) & 3.6~$\mu$m & 4.5~$\mu$m & 5.8~$\mu$m& 8.0~$\mu$m \\ \hline
590 & 1.76(0.05)\tablenotemark{a} & 1.72(0.04) & 6.33(0.27) & 21.3(0.8)\\
684 & 1.86(0.05) & 1.74(0.04) & 5.50(0.27) & 19.0(0.8)\\
758 & 2.09(0.05) & 1.58(0.04) & 5.14(0.27) & 20.8(0.8)\\ 
994 & 2.00(0.05) & 1.57(0.04) & 5.16(0.27) & 17.8(0.8)\\
\tableline
\end{tabular}
\tablenotetext{a}{Errors are shown in brackets}
\end{center}
\label{tab3}
\end{table}

%% file: tab4.tex
\begin{table}
\tablenum{4}
\begin{center}
\caption{Mid-IR fluxes of SN~2002hh - aligned, subtracted images.}
\begin{tabular}{ccccc}
\tableline\tableline
Epoch& \multicolumn{4}{c}{Flux (mJy)\tablenotemark{a}} \\ 
(d) & 3.6~$\mu$m & 4.5~$\mu$m  & 5.8~$\mu$m & 8.0~$\mu$m \\ \hline
590 & $<$0.2 & 0.227(0.010)\tablenotemark{b} & 0.82(0.07) & 2.30(0.09)\\
684 & $<$0.2 & 0.241(0.013) & 0.58(0.06) & 1.69(0.16)\\
758 & $<$0.2 & 0.09(0.07)   & 0.32(0.12) & 1.66(0.04)\\ 
\tableline
\end{tabular}
\tablenotetext{a}{The fluxes show the {\it change} in flux relative to the 994~d values.}
\tablenotetext{b}{Errors are shown in brackets}
\end{center}
\label{tab4}
\end{table}